\documentclass{elsart}

\usepackage{epsfig}

\usepackage{amssymb}

\begin{document}

\begin{frontmatter}

\title{Energy densities and equilibration in heavy ion 
collisions at $\sqrt{s_{NN}}=200~{\rm GeV}$ with the quark-gluon string model}

\author{J. Bleibel, G. Burau, Amand Faessler, C. Fuchs}

\address{Institute for Theoretical Physics, University of T\"ubingen, 
Auf der Morgenstelle 14, D-72076 T\"ubingen, Germany}

\begin{abstract}

We study thermodynamic characteristica of 
ultra-relativistic Au+Au collisions at RHIC energy 
$\sqrt{s_{NN}}=200~{\rm GeV}$ 
within the framework of a microscopic transport model, namely the 
quark-gluon string model (QGSM). The temporal evolution of 
the local energy density, transverse and longitudinal pressure and 
equilibration times are considered.  
In contrast to complete equilibration 
which is even in central reactions hardly achieved, pre-equilibrium 
stages with energy densities well above the critical energy 
density predicted by lattice QCD are established at short time scales. 
Corresponding energy density profiles at proper time $\tau =1~{\rm fm/c}$ 
compare well with hydrodynamical assumptions for initial energy density 
distributions.

\end{abstract}

\begin{keyword}
ultra-relativistic heavy ion collisions \sep 
microscopic transport model \sep 
energy density \sep 
equilibration 
\PACS 25.75.-q \sep 25.75.Nq \sep 24.10.Lx \sep 12.40.Nn
\end{keyword}
\end{frontmatter}


\section{Introduction}
\label{intro}

Ultra-relativistic heavy ion collisions as performed within experiments 
at the Relativistic Heavy Ion Collider (RHIC) in Brookhaven offer a 
unique opportunity to study the nuclear phase diagram at high
temperatures and densities \cite{Quark_matter2004}. For such
extreme conditions the fundamental theory of strong interactions
(Quantum Chromo Dynamics - QCD), i.e. lattice QCD calculations 
\cite{fodor03,allton03,Forcrand03} predict a transition to a 
plasma of deconfined quarks and gluons (QGP). This state is believed to 
behave like a strongly coupled liquid rather than a free parton 
gas \cite{shuryak04}.

It is a generally accepted opinion that the preconditions for 
the phase transition to a QGP are readily fulfilled at RHIC 
energies. In the meantime exist several experimental evidences 
that such a phase transition has indeed been seen 
\cite{Arsene:2004fa,Back:2004je,Adams:2005dq,Adcox:2004mh}.

The strongest theoretical argument comes from lattice QCD which 
predicts the phase transition at a critical temperature of around 
$170~{\rm MeV}$ which corresponds to a critical energy density of 
around $0.7 \dots 1.3~{\rm GeV/fm^3}$ (depending on the number of 
degenerate quark flavours) at zero net baryon density 
\cite{Karsch:2003jg,Laermann:2003cv}. 
Estimates from pseudorapidity distributions of charged particles 
and elliptic flow results from the PHOBOS Collaboration indicate 
that in central collisions at $\sqrt{s_{NN}}=200~{\rm GeV}$ energy 
densities $\epsilon \ge 3~{\rm GeV/fm^3}$ are produced \cite{Back:2004je} 
when the system reaches approximate equilibrium. 
Such high energy densities, which are about twenty times the energy 
density inside nuclei and about six times the energy density inside 
nucleons, exceed clearly the values for the critical energy 
density predicted from lattice QCD. 
However, lattice calculations are performed for equilibrated matter and 
infinite systems. This rises questions for the size of the space-time 
volume in which a heavy ion collision exceeds the critical energy 
density and also the degree of equilibration which is reached inside 
this volume. Therefore the
evolution and absolute magnitudes of the energy density within the
first few fm/c after the collision are of particular interest.

Also the fact that the system enters the hydrodynamical regime at 
RHIC energies $\sqrt{s_{NN}}=130\div 200~{\rm GeV}$ is often interpreted 
as an evidence for a phase transition \cite{HuSh95}. In contrast to SPS, 
at RHIC a simultanous description of elliptic flow pattern as well as of low 
$p_T$ ($p_T \lesssim 2~{\rm GeV}$) single particle spectra around mid-rapidity 
can be achieved within the hydrodynamics approach. This success has led 
to the conclusion that the created matter behaves like an {\it ideal fluid} 
with almost negligible viscosity \cite{gyulassy05,shuryak05} and that 
the time scale to reach such an locally equilibrated state is extremely 
short, i.e. less than $1~{\rm fm/c}$ \cite{heinz}.

From an experimental point of view, it is almost impossible to extract 
information about initial pressure or energy densities and their 
equilibration times. 
Hydrodynamical calculations themselves provide only little insight in 
these questions since they are  based on the assumption of local thermal 
equilibrium and short equilibration times. However, also hydrodynamics 
does not provide a perfect description of RHIC data. In particular, 
deviations of the hydrodynamical predictions for $v_2$ and $v_4$ 
from the measured elliptic flow were recently interpreted 
as a breakdown of the hydrodynamical picture and a hint towards incomplete 
equilibration \cite{bahl05}. 
In the hydrodynamics approach an ansatz for the early stage, i.e. 
at about $1~{\rm fm/c}$ after the initial collision, is used as input for the
evolution of the system. This requires to fix the initial energy density and
net-baryon number density profiles.  
Usually for both distributions a boost-invariant
distribution over a wide range of $\eta_s$, the so called space-time
rapidity is chosen.

Transport models on the other hand, allow to study the 
dynamical evolution of thermodynamical properties of the colliding 
system from a microscopic point of view. String-cascade models of that type, 
e.g. UrQMD \cite{URQMD1,URQMD2}, HSD \cite{HSD1,HSD2} or the quark-gluon 
string model (QGSM), do not contain an explicit phase transition to a 
deconfined partonic state but they can be used to study the 
preconditions for a possible phase transition and the results can 
be compared with the assumptions made by hydrodynamical approaches. 
The quark-gluon string model, applied in 
the present work and described in detail in
Refs. \cite{QGSM0a,QGSM0b,QGSM1a,QGSM1b,QGSM1c} is a string model
based on Gribov Regge Theory. The main assumption within these kind of
(string-) models is that hadrons are produced as a result of excitation
and decay of open strings with different quarks or diquarks at their
ends. The QGSM provides a safe basis for the present investigations since 
it has been demonstrated to describe the elliptic flow bulk properties 
at SPS and in particular at RHIC energies reasonably well  
\cite{QGSM_flow1a,QGSM_flow1b,QGSM_flow2}.

\section{Evolution of energy density}
\label{edens}

In order to obtain an impression of the temporal evolution of the 
energy density $\epsilon$ we consider as two typical examples a 
central  ($b=0~{\rm fm}$) and a semi-peripheral ($b=8~{\rm fm}$) 
Au+Au reaction at top RHIC energy of $\sqrt{s_{NN}}=200~{\rm GeV}$. 
In the present approach the energy density is locally determined 
in cells of $1~{\rm fm^3}$ volume through the scalar mass density of 
all formed and pre-formed hadrons inside these cells\footnote{Exceptions 
to this specification are appropriately indicated.}. 
All QGSM results are obtained from averages over 200 central reactions 
($b=0~{\rm fm}$), respectively 600 semi-peripheral reactions ($b=8~{\rm fm}$). 
For simplicity, kinetic contributions to the energy density are neglected. 
Thus, the $\epsilon$ values presented in the following analysis have to be 
interpreted as lower limits of the total amount achieved in the 
collisions. To clarify this a bit more, let us consider a central 
collision at the very first instant when the two Lorentz contracted 
gold nuclei with a longitudinal size of $\approx 0.1~{\rm fm}$ and a 
transverse area of $\approx 150~{\rm fm^2}$ overlap totally. This yields 
an energy density of $\approx 25~{\rm GeV/fm^3}$ in this approach.

\begin{figure}[ht]
  \epsfig{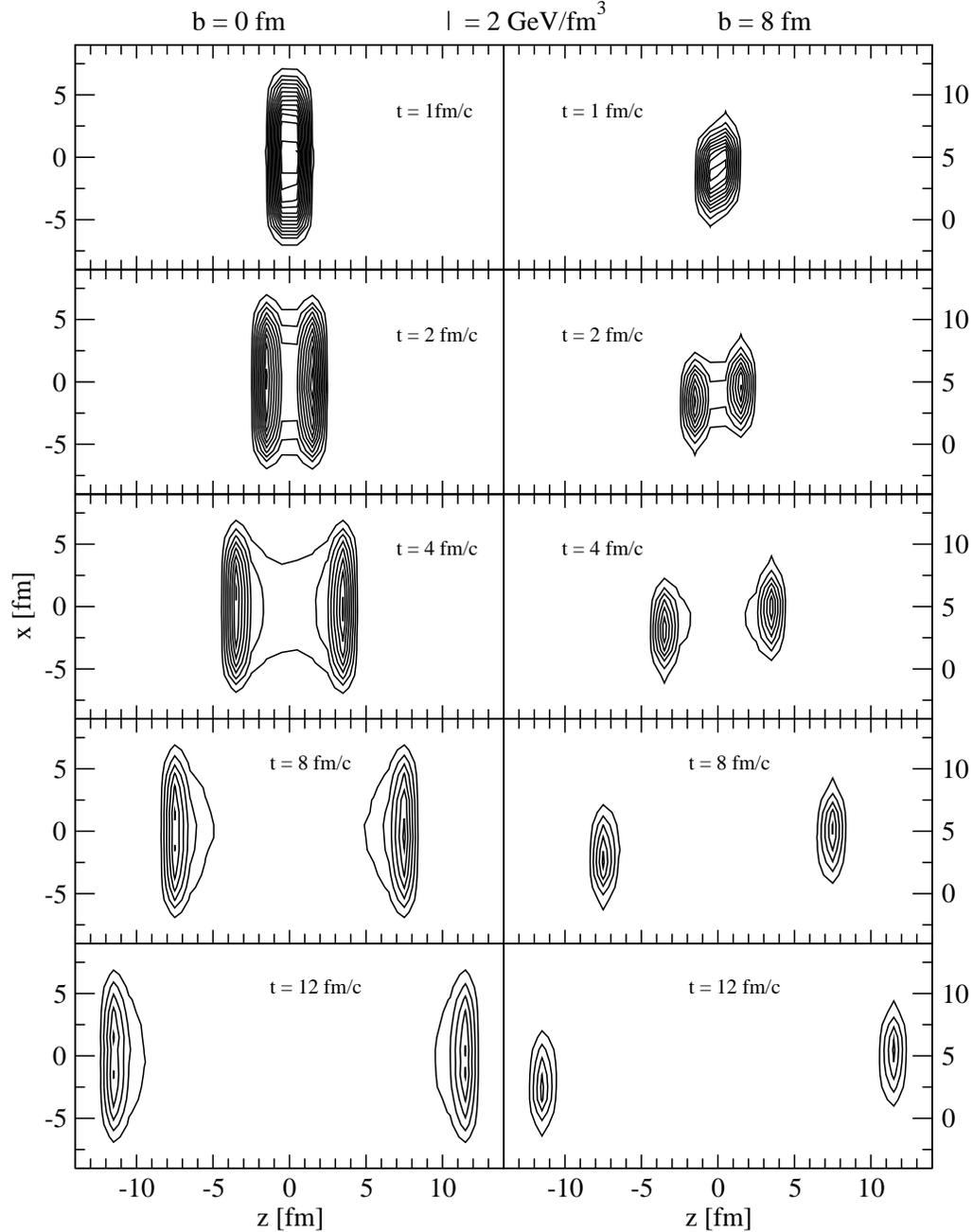}
  \caption{\label{FIG1}Evolution of the energy density in the x-z plane
  for two different impact parameters, $b=0~{\rm fm}$ (left) and 
  $b=8~{\rm fm}$ (right). Each line corresponds to an energy density 
  interval of $2~{\rm GeV/fm^3}$.}
\end{figure}

Fig. \ref{FIG1} displays the time evolution of the local energy 
density $\epsilon$ in the x-z plane as a function of the system time $t$ 
in the c.m. frame of the colliding nuclei. 
One should thereby keep in mind that the corresponding hypersurfaces 
where $\epsilon ({\bf x}, t={\rm const.})$ is evaluated are essentially 
different form hypersurfaces of proper time 
$\epsilon ({\bf x}, \tau={\rm const.})$  which are used to characterize 
the time evolution in hydrodynamics. The latter case will be discussed 
later on.

One immediately recognises in the contour plot of Fig. \ref{FIG1}, that 
most particles do not stay in the central collision zone but rather form 
some kind of ``shock waves'', i.e. two disks in the fragmentation regions 
occupied with large numbers of produced particles which propagate with high 
rapidities in front of the expanding system. As a result of that, the energy 
density in these zones stays at a high level for the whole time under 
consideration and decreases only slowly. Therefore most collisions happen 
within these zones while the center of the volume between the two disks is 
rapidly thinning out of hadrons. 
After the first initial collisions energy densities of a few tens of GeV 
are achieved in the overlap zone of the central as well as semi-peripheral 
reaction.

\begin{figure}[h]
  \vspace{0.5cm}
  \centering
  \epsfig{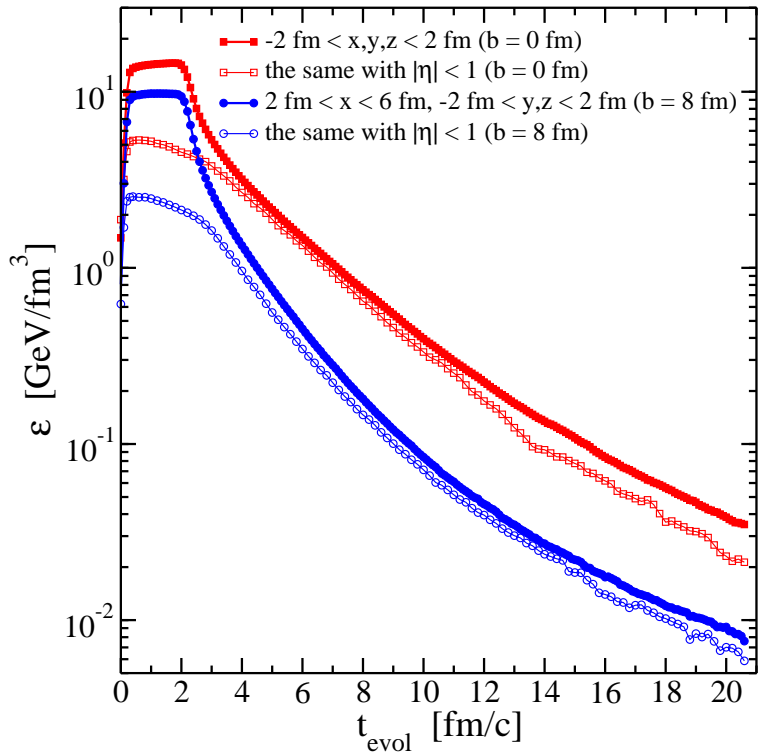}
  \caption{\label{FIG2}Evolution of the energy density for a central cell 
    in the overlap zone of the collision given by 
    $-2~{\rm fm}<x,y,z<2~{\rm fm}$ for 
    $b=0~{\rm fm}$ and $2~{\rm fm}<x<6~{\rm fm}$, 
    $-2~{\rm fm}<y,z<2~{\rm fm}$ for $b=8~{\rm fm}$. The curves with 
    open symbols are the same but only for particles with a 
    pseudorapidity of $|\eta| \le 1$. See discussion in the text.}
\end{figure}

To study the initial phase and the overlap zone more closely, 
we consider now the average energy density inside a larger 
volume of $64~{\rm fm^3}$. For the central reaction we choose 
the central reaction zone, i.e. the most central cell given by 
$-2~{\rm fm}<x,y,z<2~{\rm fm}$ and for the  $b=8~{\rm fm}$ case 
a volume shifted by $4~{\rm fm}$ in x-direction, 
i.e. $2~{\rm fm}<x<6~{\rm fm}$, $-2~{\rm fm}<y,z<2~{\rm fm}$. 
Fig. \ref{FIG2} displays the time evolution of $\epsilon$ in the 
corresponding cells. One sees, that for the central collision a 
maximum energy density of around $15~{\rm GeV/fm^3}$ is reached 
while the semi-peripheral collision reaches a maximal $\epsilon$ 
of around $10~{\rm GeV/fm^3}$. Again, we want to emphasize that 
all particles are taken into account, i.e. first of all no rapidity 
cut is applied. 
The energy density stays almost constant during the first $2~{\rm fm/c}$. 
It is even  slightly increasing due to the ongoing particle production 
through scattering of projectile and target nucleons and the 
corresponding continous energy deposit. 
As soon as the fastest particles start to leave the cell, what happens 
at around $t\approx 2~{\rm fm/c}$, $\epsilon$ is rapidly decreasing. 
This means that we obtain energy densities of $15~{\rm GeV/fm^3}$ 
in a space-time volume of at least $\sim 130~{\rm fm}^4$ in the 
central collision and correspondingly about $10~{\rm GeV/fm^3}$ in the 
same volume for the  semi-peripheral reaction. However, at that time 
the system is far from local equilibrium, as will be seen later on. 
It is stated in Ref. \cite{Back:2004je} that the value of the energy 
density for such a system, although well defined, may not be very 
interesting. The potentially more interesting quantity should be the 
energy density carried by particles which are closer to equilibrium 
conditions, wherefore the analysis in \cite{Back:2004je} was 
restricted to hadrons with pseudorapidities $|\eta| \le 1$. 
To make a comparison between the estimates of the PHOBOS Collaboration 
and the QGSM results more transparent, two additional curves for the 
temporal evolution of $\epsilon$ with the same $|\eta| \le 1$ restriction 
are depicted in Fig. \ref{FIG2}. 
Here, energy densities in the order of $\ge 4~{\rm GeV~fm}^{-3}$ in the 
central collisions and $\ge 2~{\rm GeV~fm}^{-3}$ in the semi-peripheral 
reactions are obtained after $2~{\rm fm/c}$ time of evolution. 
These values are in nice agreement with the lower limit estimation of 
$\epsilon \ge 3~{\rm GeV~fm}^{-3}$ produced in central Au+Au collisions 
at RHIC after $1-2~{\rm fm/c}$, as it is reported by the PHOBOS 
Collaboration \cite{Back:2004je}. Note, that also the QGSM results are 
lower limits of the produced energy densities as mentioned before. 
The actual densities could easily be significantly larger. 
Nevertheless, the parton-string cascade model is not able to equilibrate 
completely the system at such short time scales, as we will discuss later on.

\begin{figure}[ht]
  \psfig{file=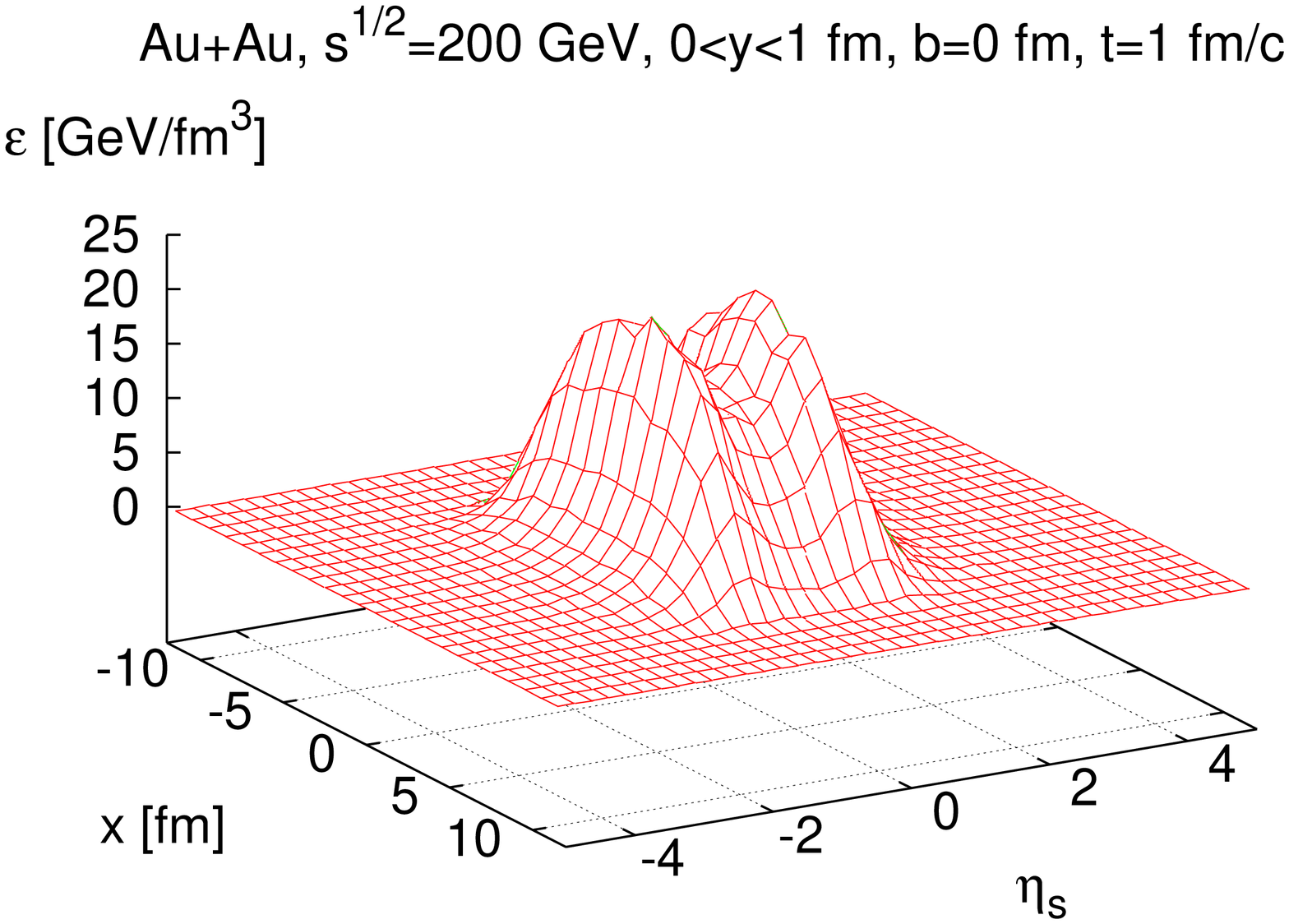,height=6.8cm,angle=-90}
  \psfig{file=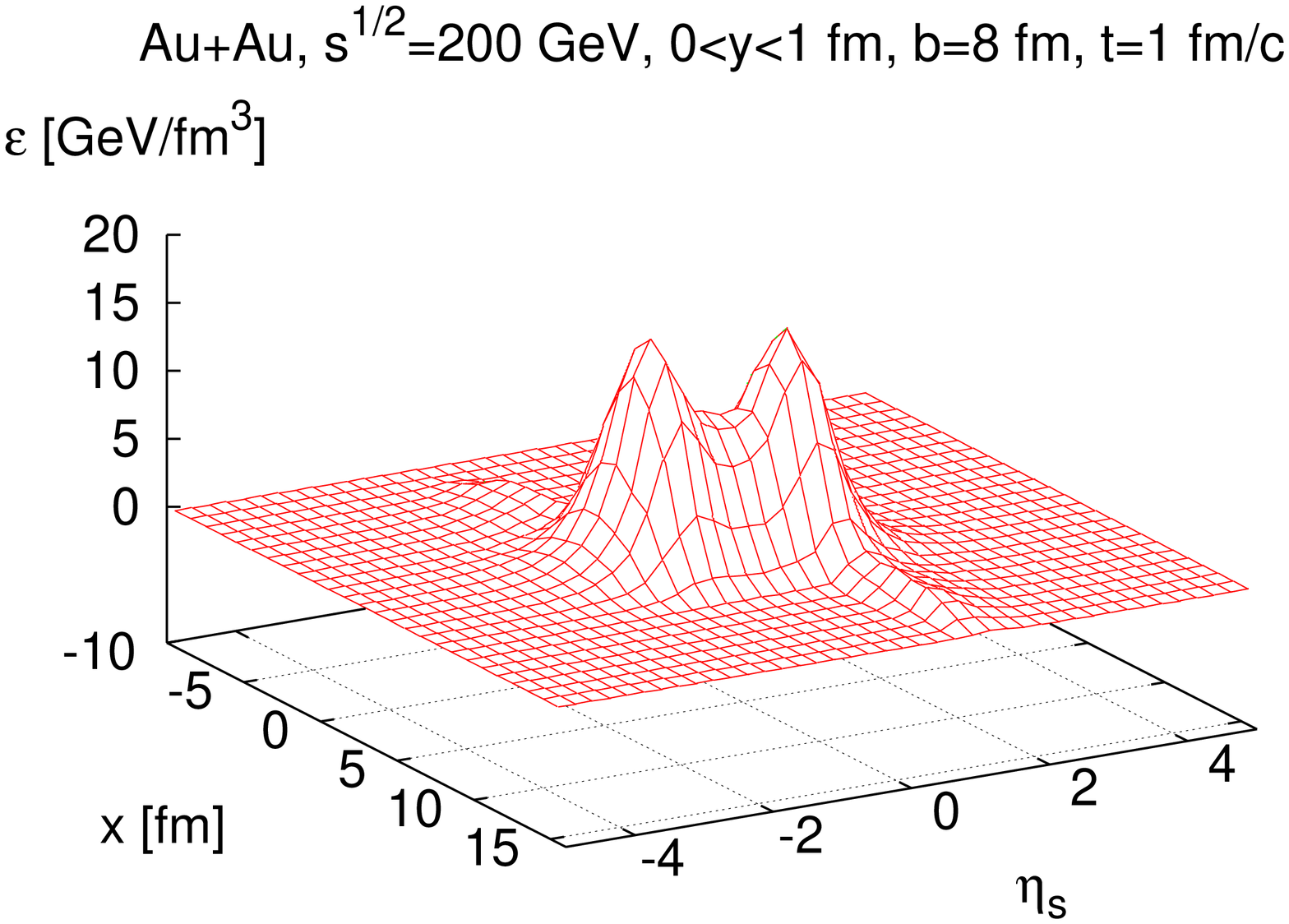,height=6.8cm,angle=-90}
  \psfig{file=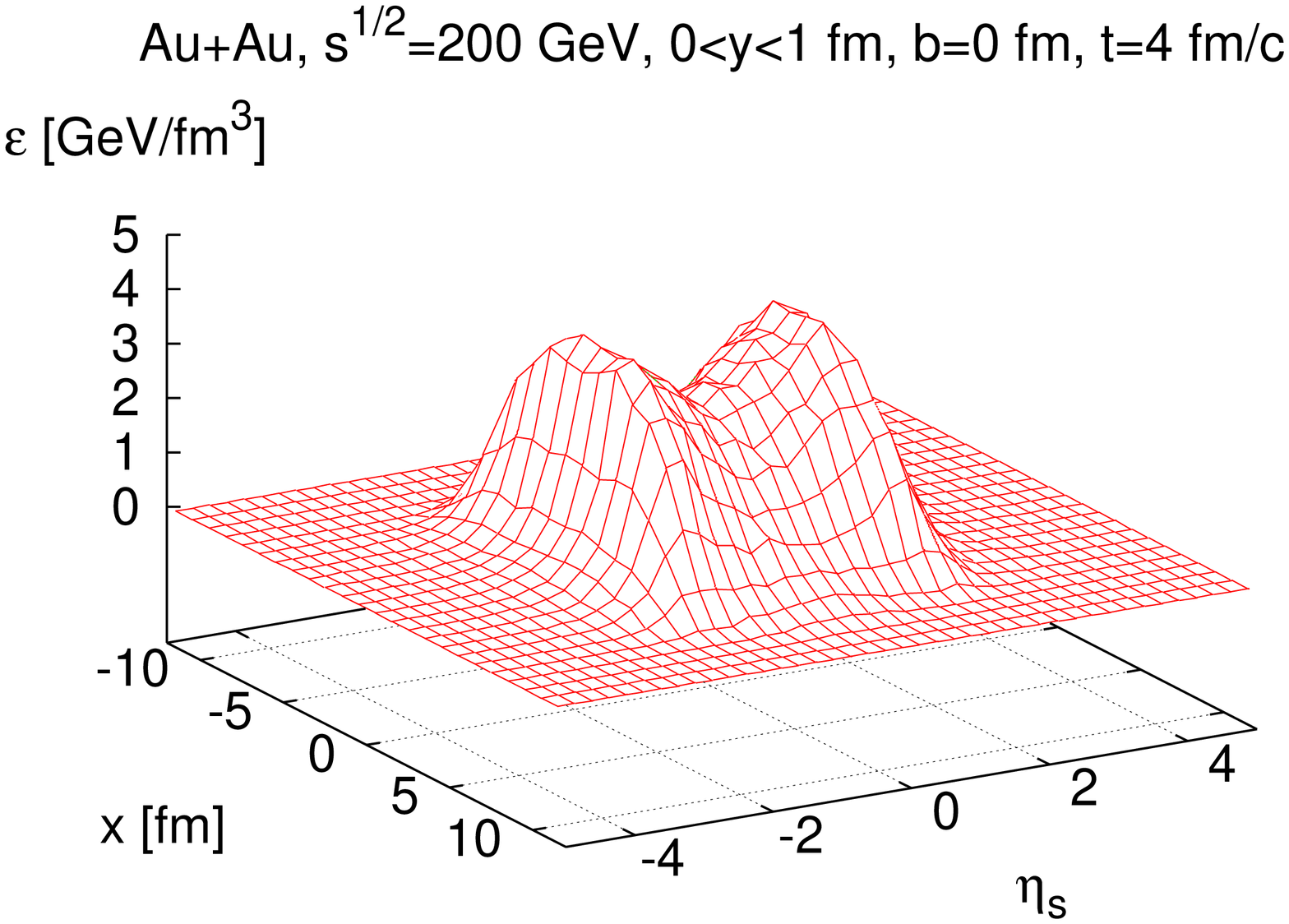,height=6.8cm,angle=-90}
  \psfig{file=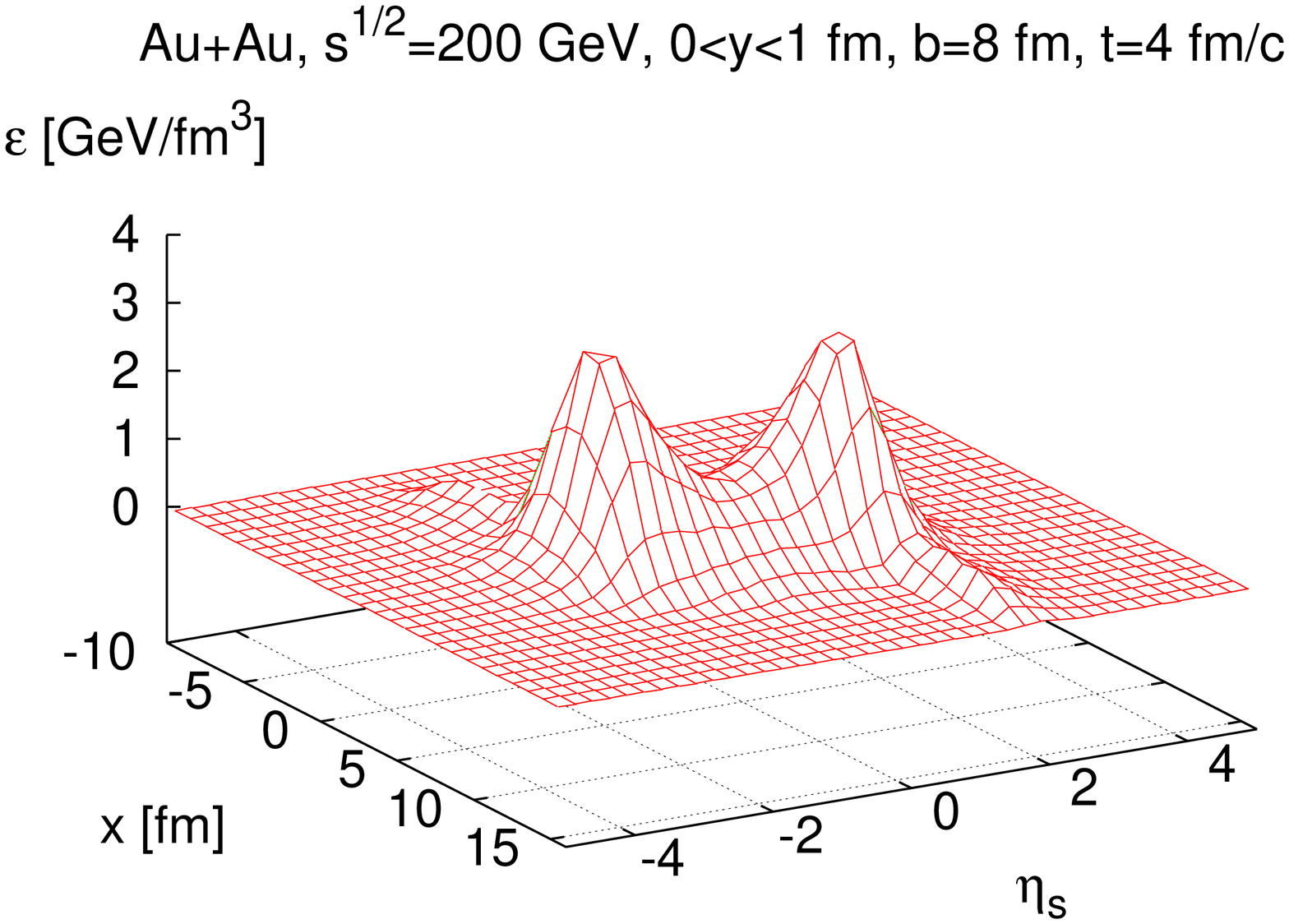,height=6.8cm,angle=-90}
  \psfig{file=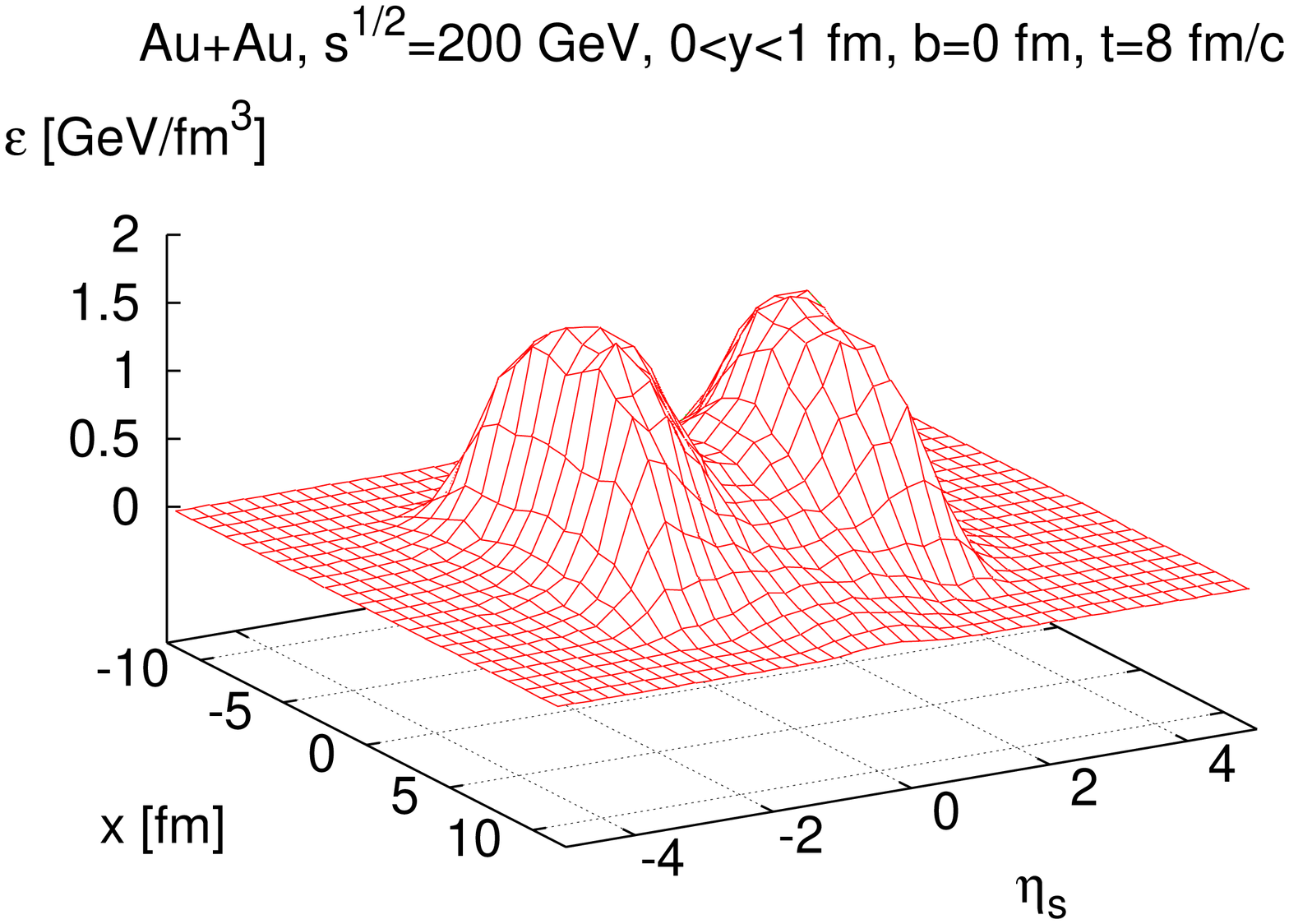,height=6.8cm,angle=-90}
  \psfig{file=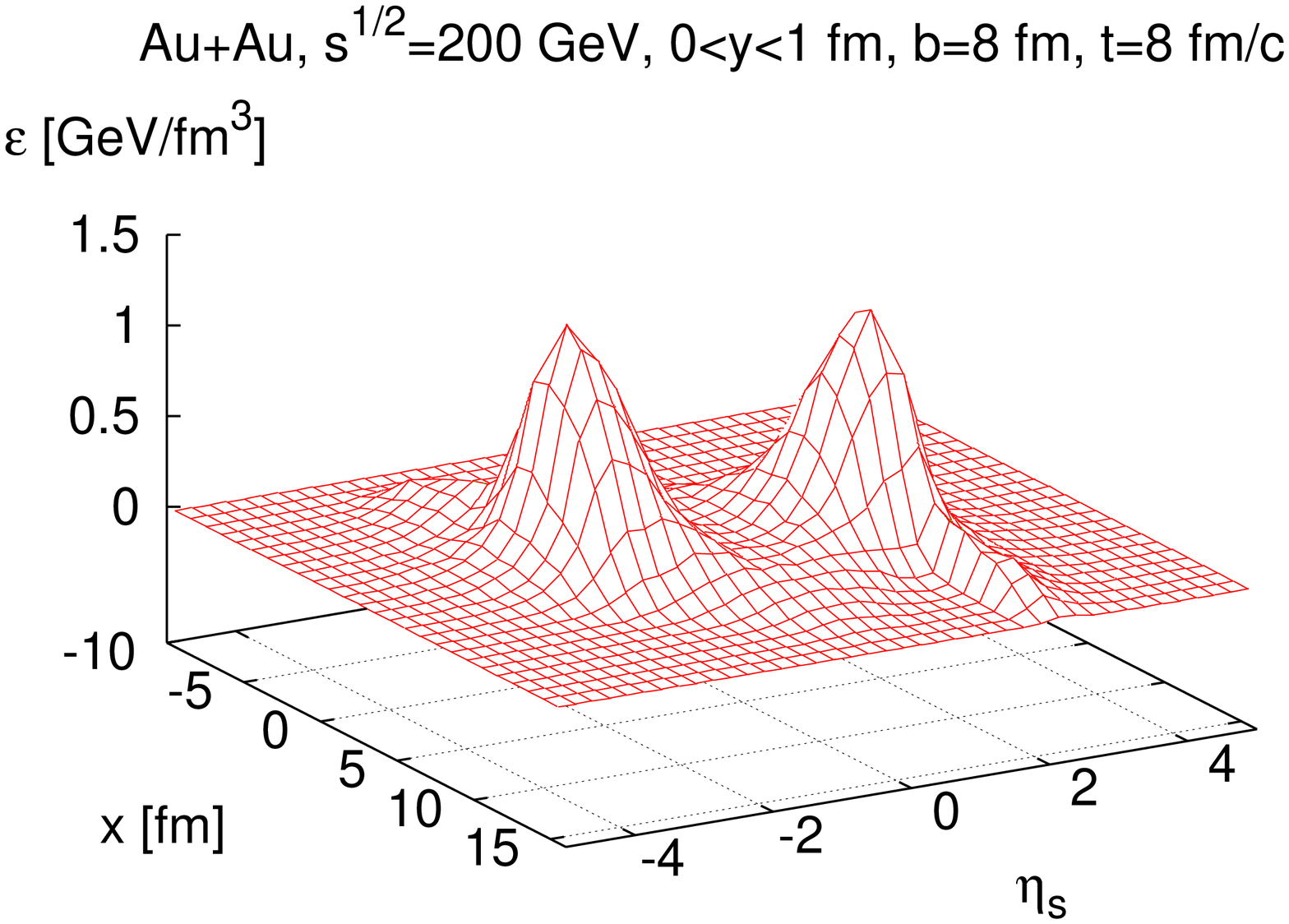,height=6.8cm,angle=-90}
  \psfig{file=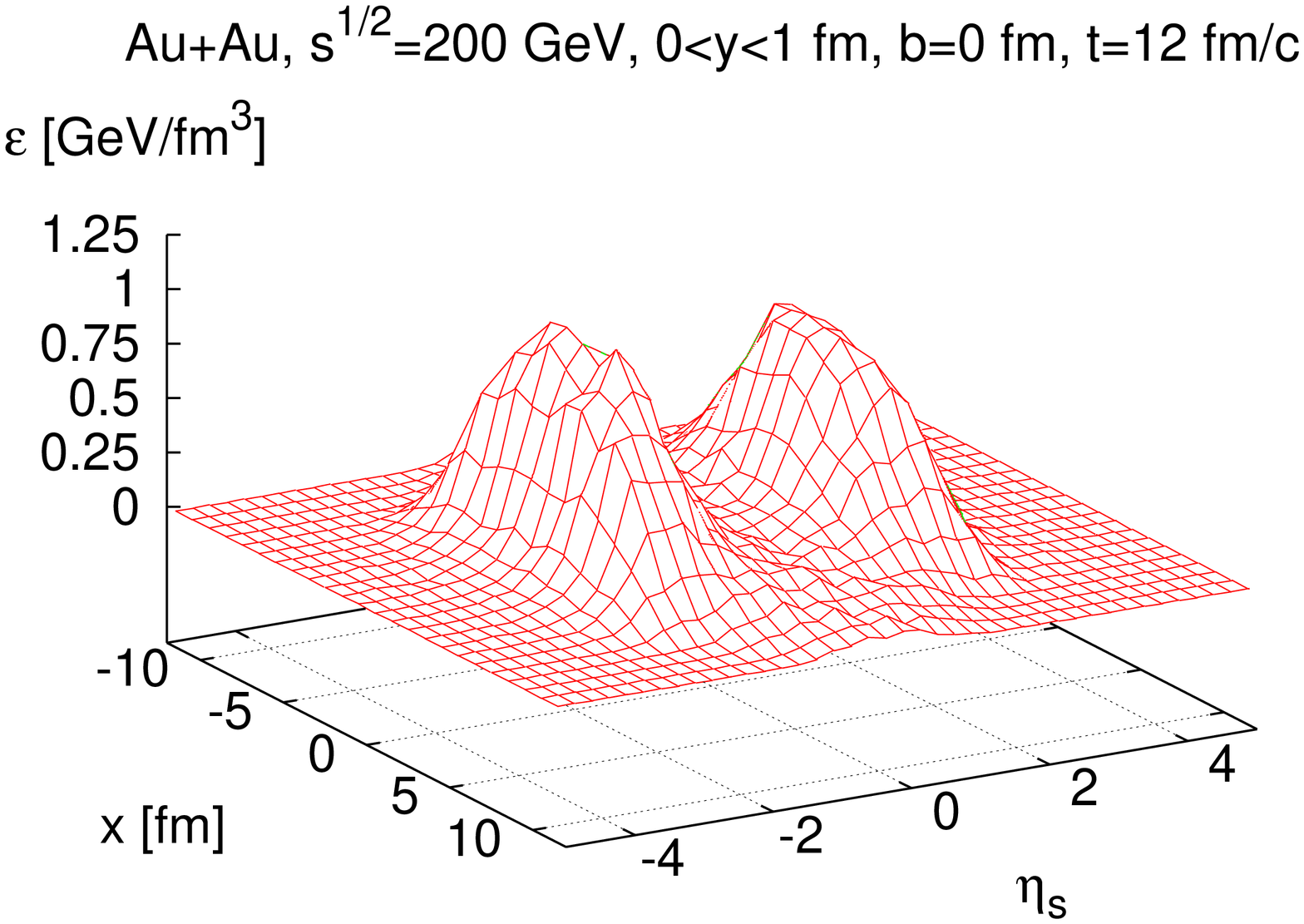,height=6.8cm,angle=-90}
  \psfig{file=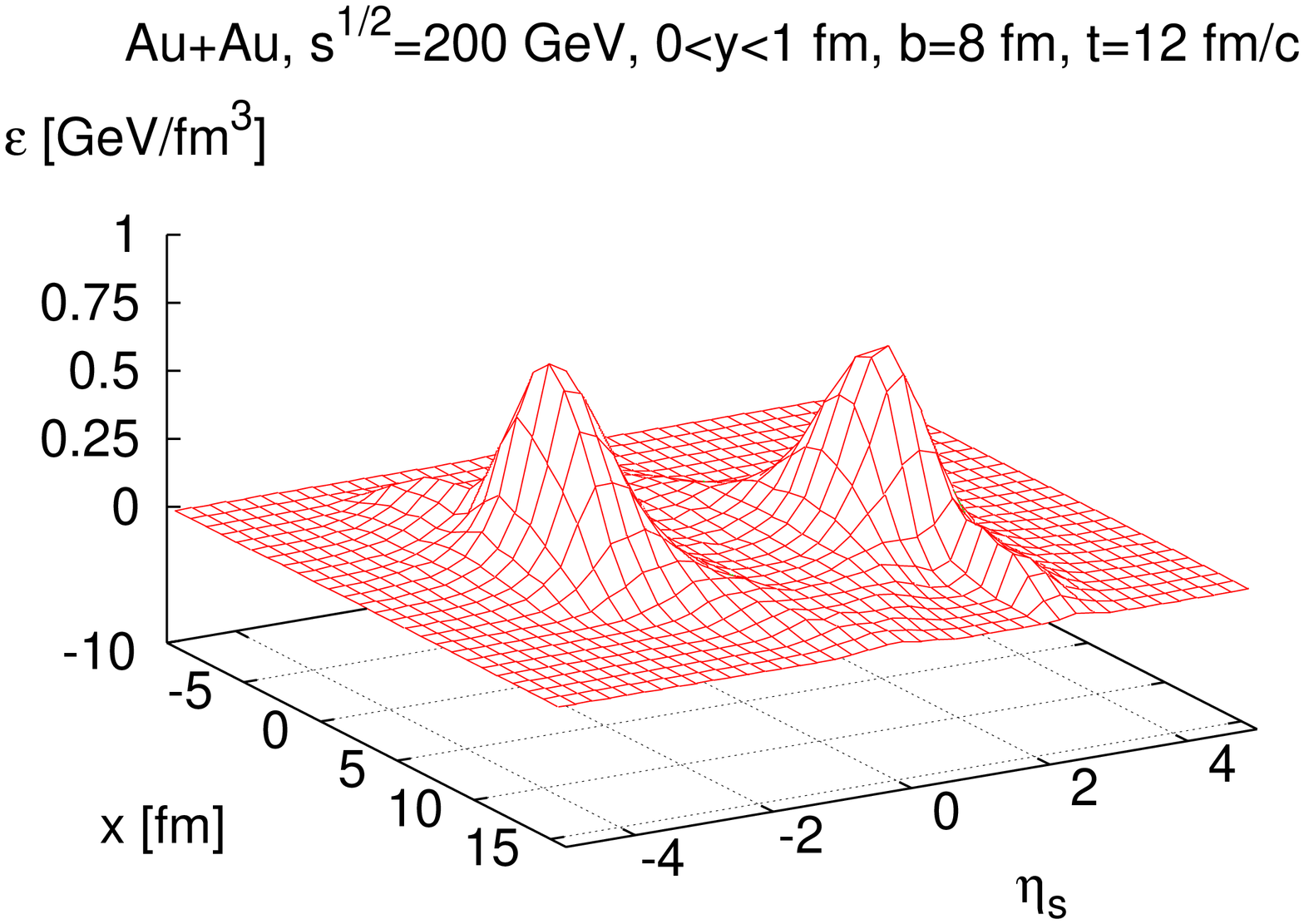,height=6.8cm,angle=-90}
  \caption{\label{E_1-12fm}Snapshots of the energy density $\epsilon$ 
  in the $x-\eta_s$ plane at times $t=1,~4,~8,~12~{\rm fm/c}$. Note the 
  different scales for $\epsilon$.}
\end{figure}

As clearly seen from Fig. \ref{FIG1}, the geometry of the colliding 
system is dominated by the leading particles which move practically 
on the light cone in beam $(z)$ direction. In order to remove distortions 
from Lorentz contraction effects in a rapidly longitudinally 
expanding system one can introduce the so called space-time rapidity 
$\eta_s$. This is done by the introduction of cylindrical coordinates 
in $t$- and $z$-direction according to
\begin{equation}
(t,x,y,z)=(\tau\cosh{\eta_s},x,y,\tau\sinh{\eta_s})
\end{equation}
with the longitudinal proper time $\tau=\sqrt{t^2-z^2}$. 
The space-time rapidity is then given by the following expression:
\begin{equation}
  \eta_s=\frac{1}{2}\,\ln{\frac{t+z}{t-z}}~~.
\end{equation}
The evolution of the energy density can now be displayed as a function
of the boost-invariant variable $\eta_s$. Fig. \ref{E_1-12fm} shows 
the energy density in the $x-\eta_s$ plane at the same time steps as 
have been used in Fig. \ref{FIG1}. The $y$-coordinate is thereby 
restricted to the interval $0<y<1~{\rm fm}$. 
Also as a function of $\eta_s$ the system has an energy density distribution 
with a nontrivial shape in the reaction plane. For the central collision, 
$\epsilon$ is rather Gaussian distributed in x-direction whereas it shows 
a characteristic double-peaked structure in $\eta_s$-direction, where the 
maxima follow the leading particles. This qualitative feature of the 
energy density distribution is practically maintained in the overlap zone 
of the semi-peripheral collision, albeit it is narrower distributed 
along the x-axis. Additionally, one clearly recognises the $\epsilon$ 
contributions of the spectators (``shoulders'' at large $|x|$ and $|\eta_s|$) 
which propagate with about the same velocity as the high density bumps formed 
by the leading particles. 
Within the first few fm/c of the expansion the central values are 
large, i.e. $\epsilon\approx 10\div 15~{\rm GeV/fm^3}$ and the 
distributions are more compact. However, as already seen from Fig. \ref{FIG2}, 
$\epsilon$ drops rapidly when the system expands. This decrease with 
increasing time is more pronounced in the central zone forming the 
``dip'' than for the bumps following the leading particles. But also 
there the absolute values drop significantly and the peaks get washed 
out due to ongoing secondary collisions between the produced hadrons.

\begin{figure}[ht]
  \psfig{file=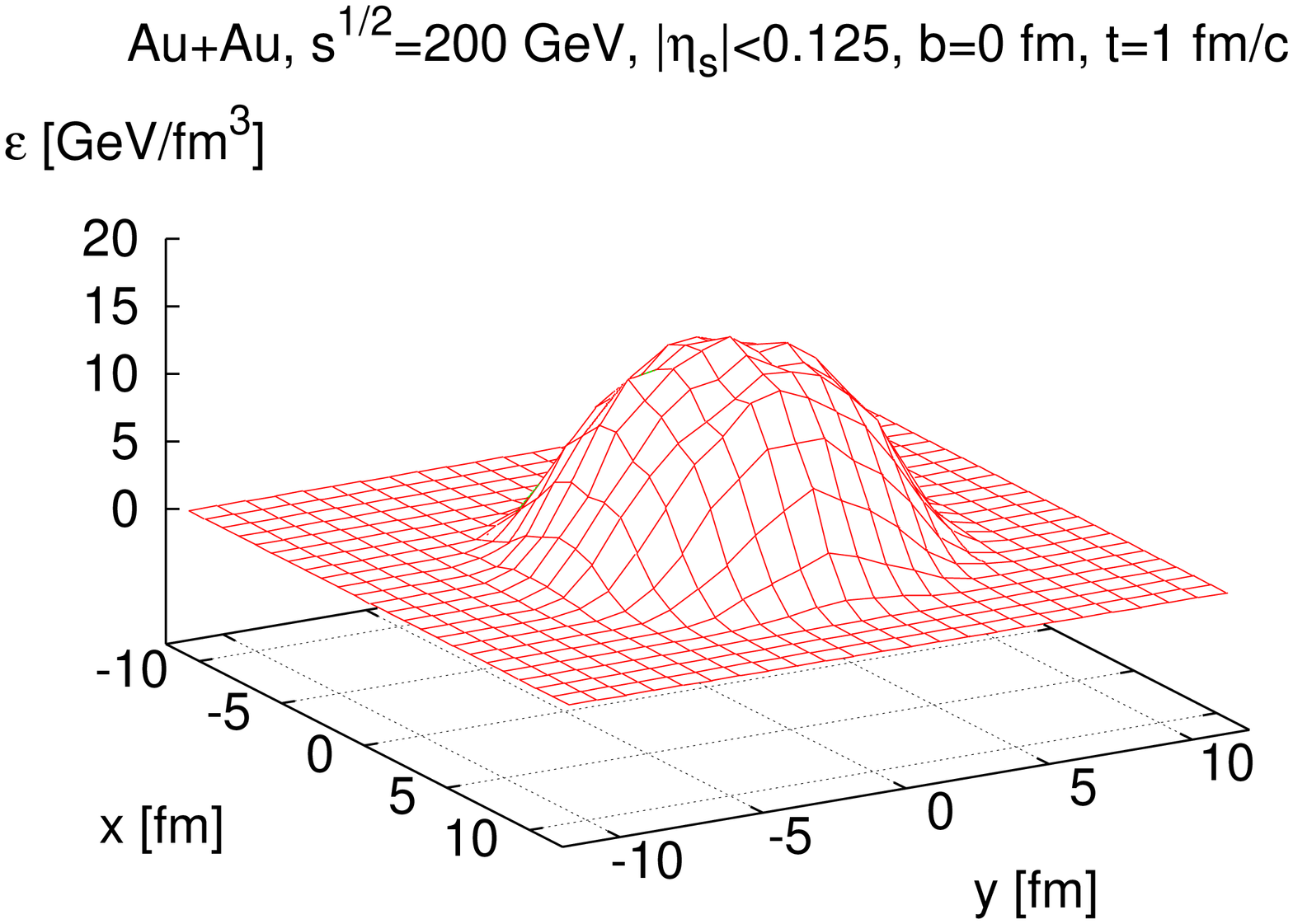,height=6.8cm,angle=-90}
  \psfig{file=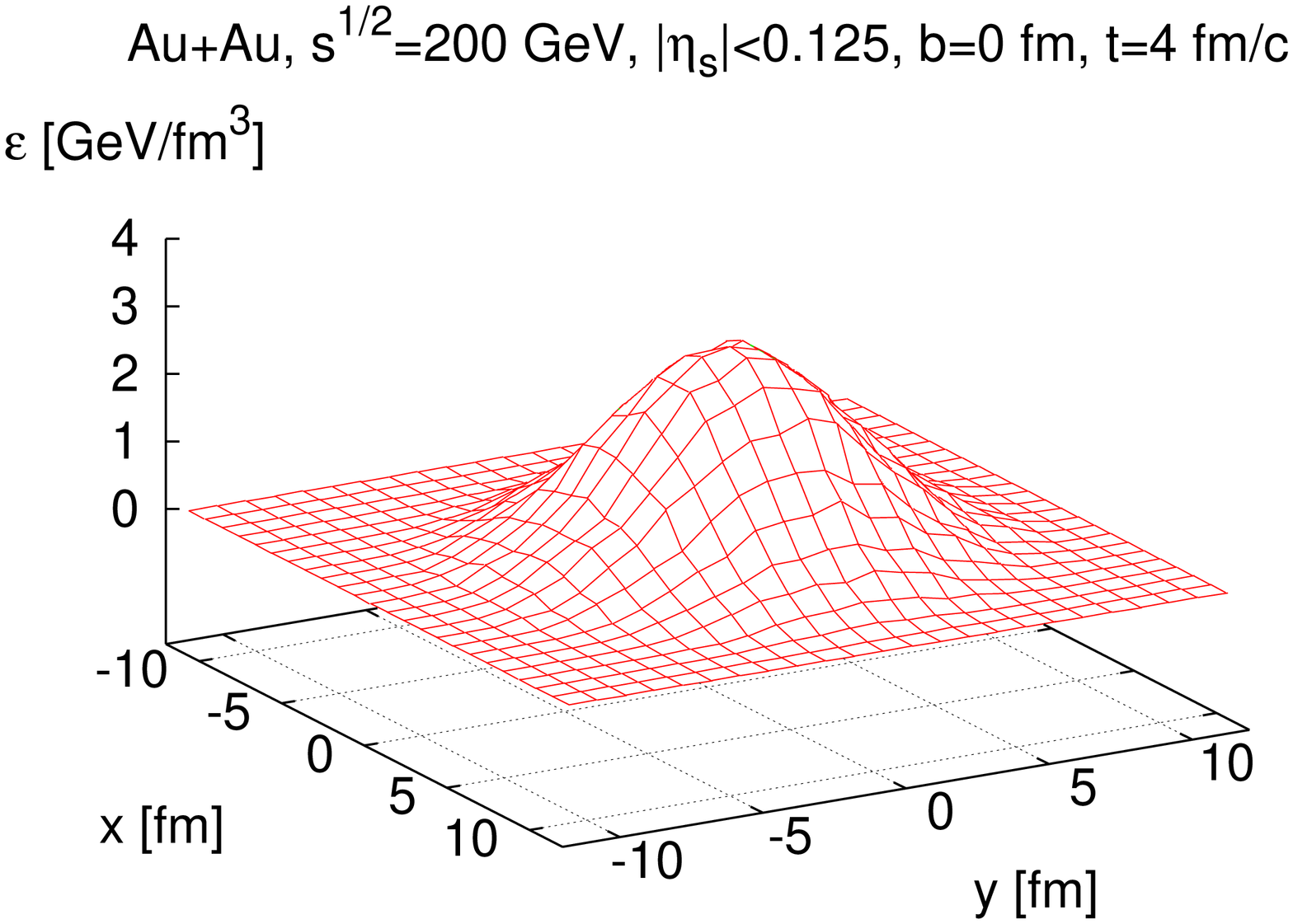,height=6.8cm,angle=-90}
  \psfig{file=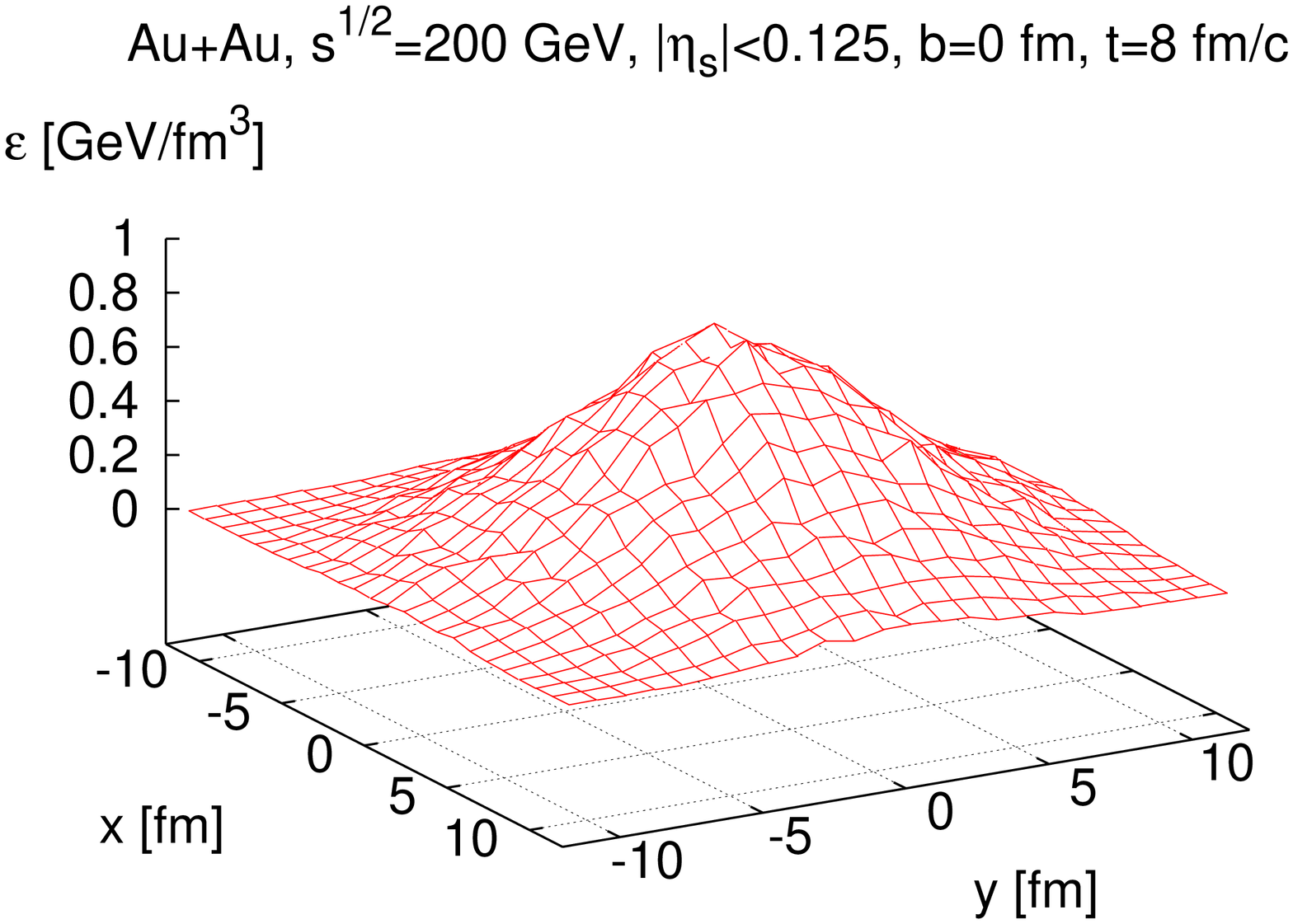,height=6.8cm,angle=-90}
  \psfig{file=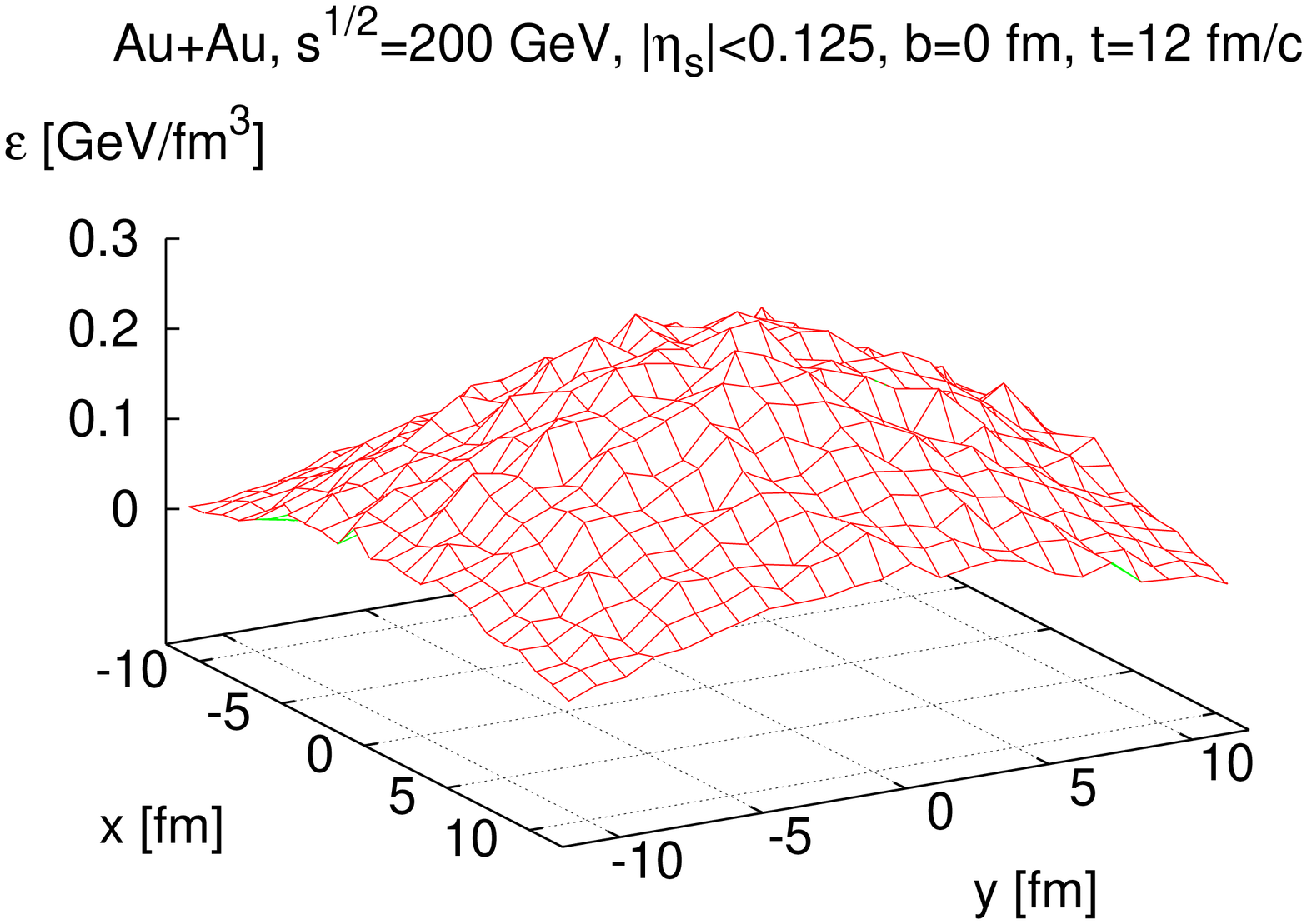,height=6.8cm,angle=-90}
  \caption{\label{E_eta0} Snapshots of the energy density in the
  transversal plane for a space-time rapidity slice of 
  $|\eta_s| < 0.125$ obtained for a central ($b = 0~{\rm fm}$) 
  Au+Au collision at times $t=1,~4,~8,~{\rm and}~12~{\rm fm/c}$.}
\end{figure}

Complementary to the energy density in the reaction plane, 
the corresponding distributions of $\epsilon$ in the transversal (x-y) 
plane are depicted in Fig. \ref{E_eta0}. Here we consider only the central 
reaction with $b = 0~{\rm fm}$. 
The contributions of the leading particles are subtracted by restricting 
the space-time rapidity to the interval $|\eta_s| < 0.125$ in all the 
four time steps under consideration. The distributions look now more 
isotropic, i.e. Gaussian shaped in x- and y-direction, and are peaked at 
the origin of the transversal plane. As already seen from the previous 
figures, the peak value of $\epsilon$ reaches up to $15~{\rm GeV/fm^3}$ 
for $t=1~{\rm fm/c}$. Although the energy density is rapidly decreasing 
in the transverse directions as well as in time, there exist definitely 
regions with values of $\epsilon$ within the first $8~{\rm fm/c}$, 
that are higher than the expected critical energy density for a 
phase transition to a deconfined phase.

\begin{figure}[ht]
  \psfig{file=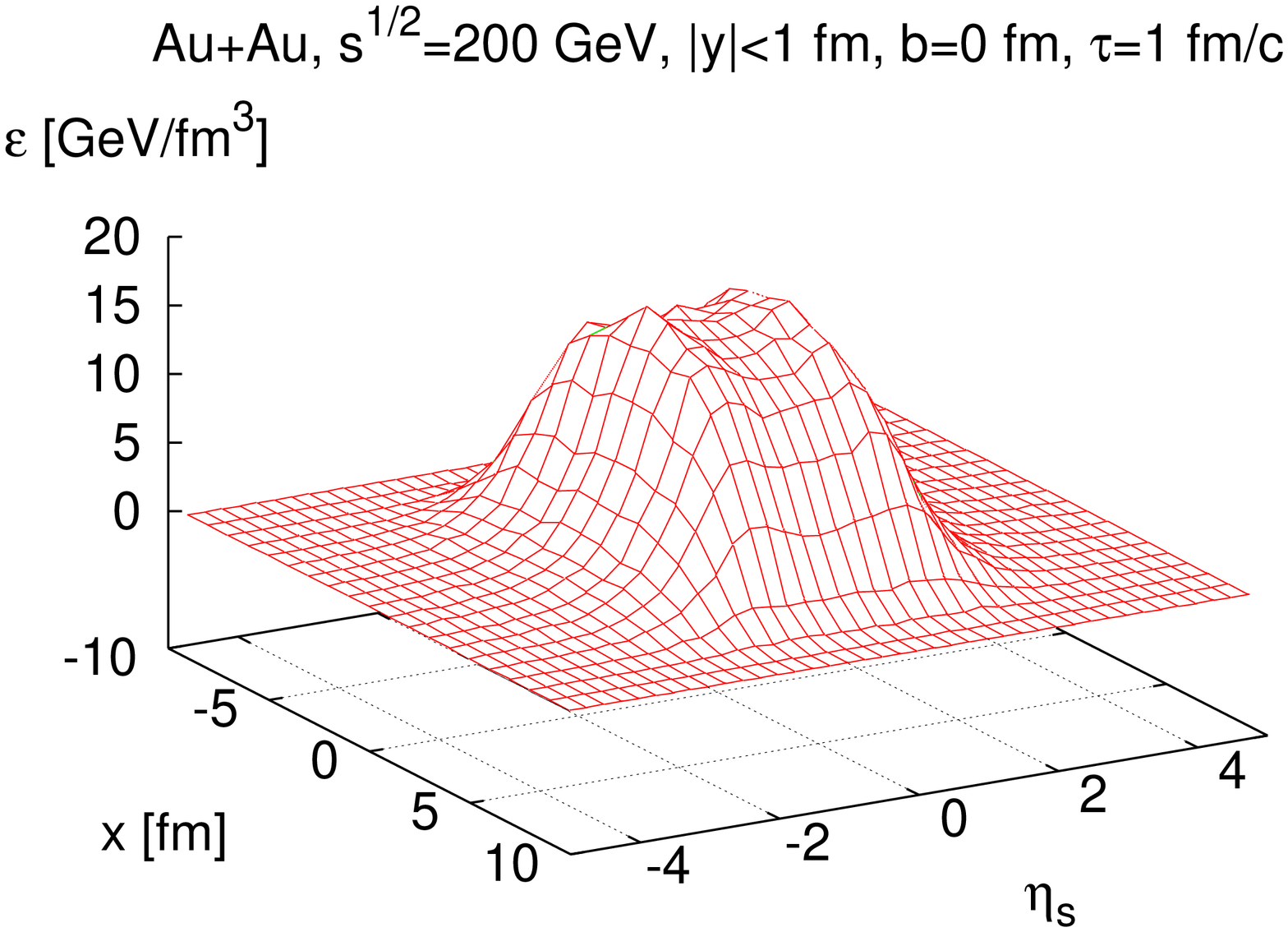,height=6.8cm,angle=-90}
  \psfig{file=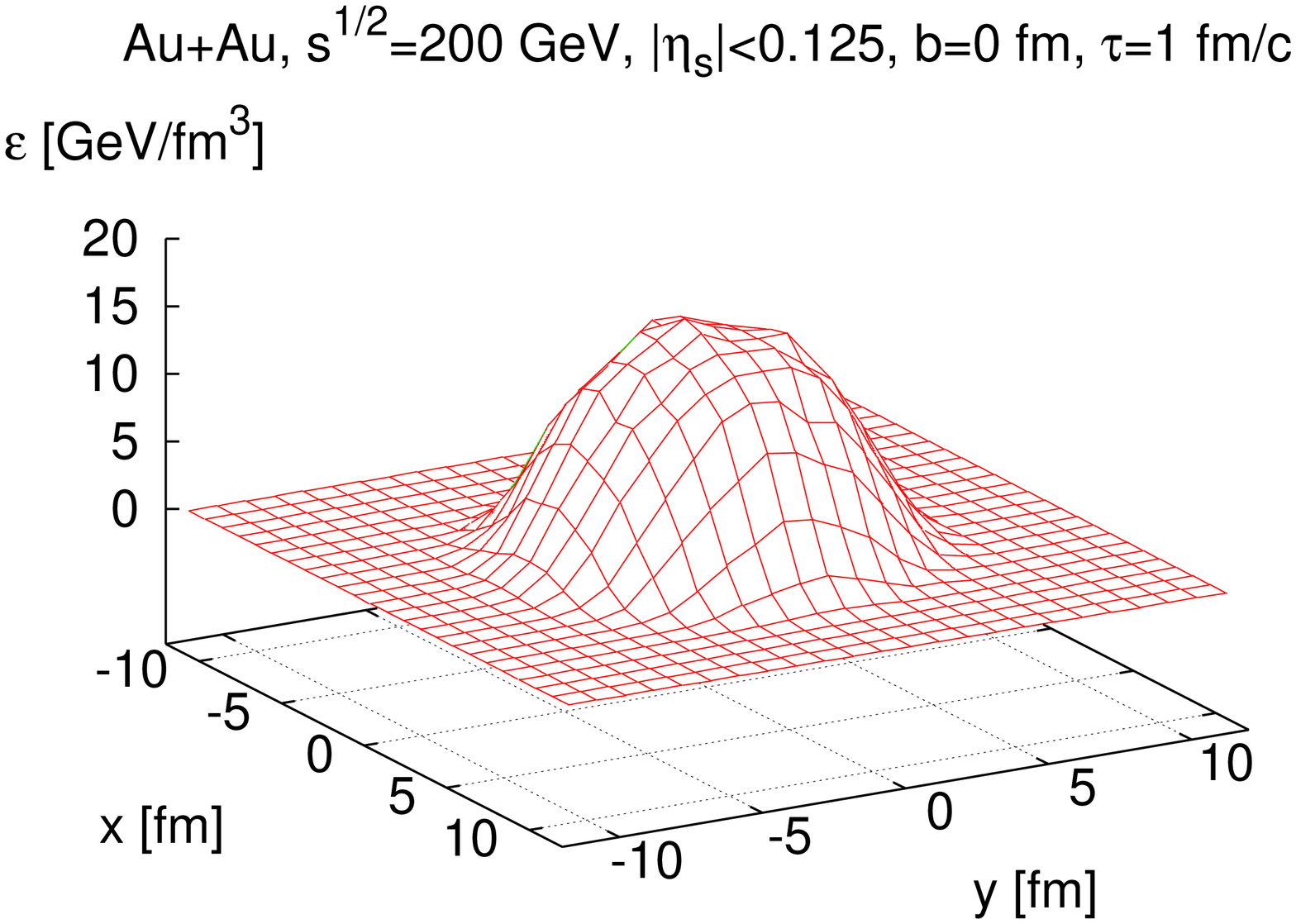,height=6.8cm,angle=-90}
  \caption{\label{E_tau1} Energy density of central Au+Au collisions in the 
  reaction plane ($|y| < 1~{\rm fm}$ and $|x|,~|\eta_s| \ge 0$) -- left panel, 
  and in the transversal plane ($|\eta_s| < 0.125$ and $|x|,~|y| \ge 0$) -- 
  right panel, evaluated for constant longitudinal proper time 
  $\tau=1~{\rm fm/c}$.}
\end{figure}

Finally, the energy density $\epsilon$ is evaluated for the central 
reaction at constant proper time $\tau = 1~{\rm fm/c}$. The QGSM result 
is shown in Fig. \ref{E_tau1}. The depicted distributions can be 
directly compared with initial energy density distributions used 
in hydrodynamical calculations \cite{hydro,hydro1,hydro2}. 
The $\epsilon$ distribution in the reaction plane as obtained in the 
QGSM is rather similar to the initial energy densities investigated by 
Hirano et al. for Au+Au collisions at $\sqrt{s_{NN}} = 130~{\rm GeV}$ 
in a hydrodynamical model \cite{hydro,hydro1}. The distribution 
is nearly flat in the region $|\eta_s| \le 1.0$ and connects to vacuum 
as a kind of Gaussian function sharply decreasing in the forward and 
backward space-time rapidity regions. 
\begin{figure}[ht]
  \begin{minipage}{7cm}
  \psfig{file=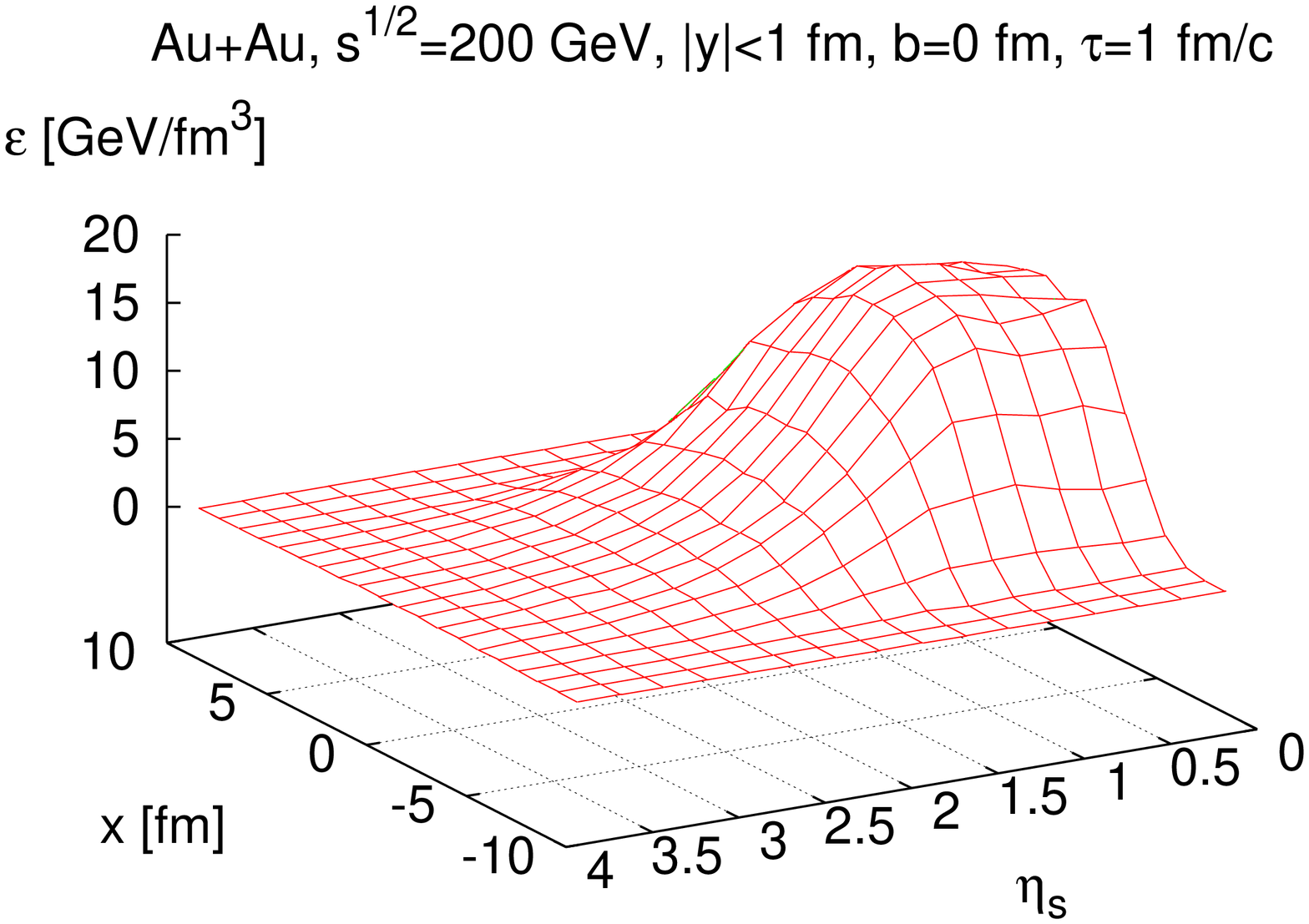,height=6.8cm,angle=-90}
  \end{minipage}
  \begin{minipage}{7cm}
  \psfig{file=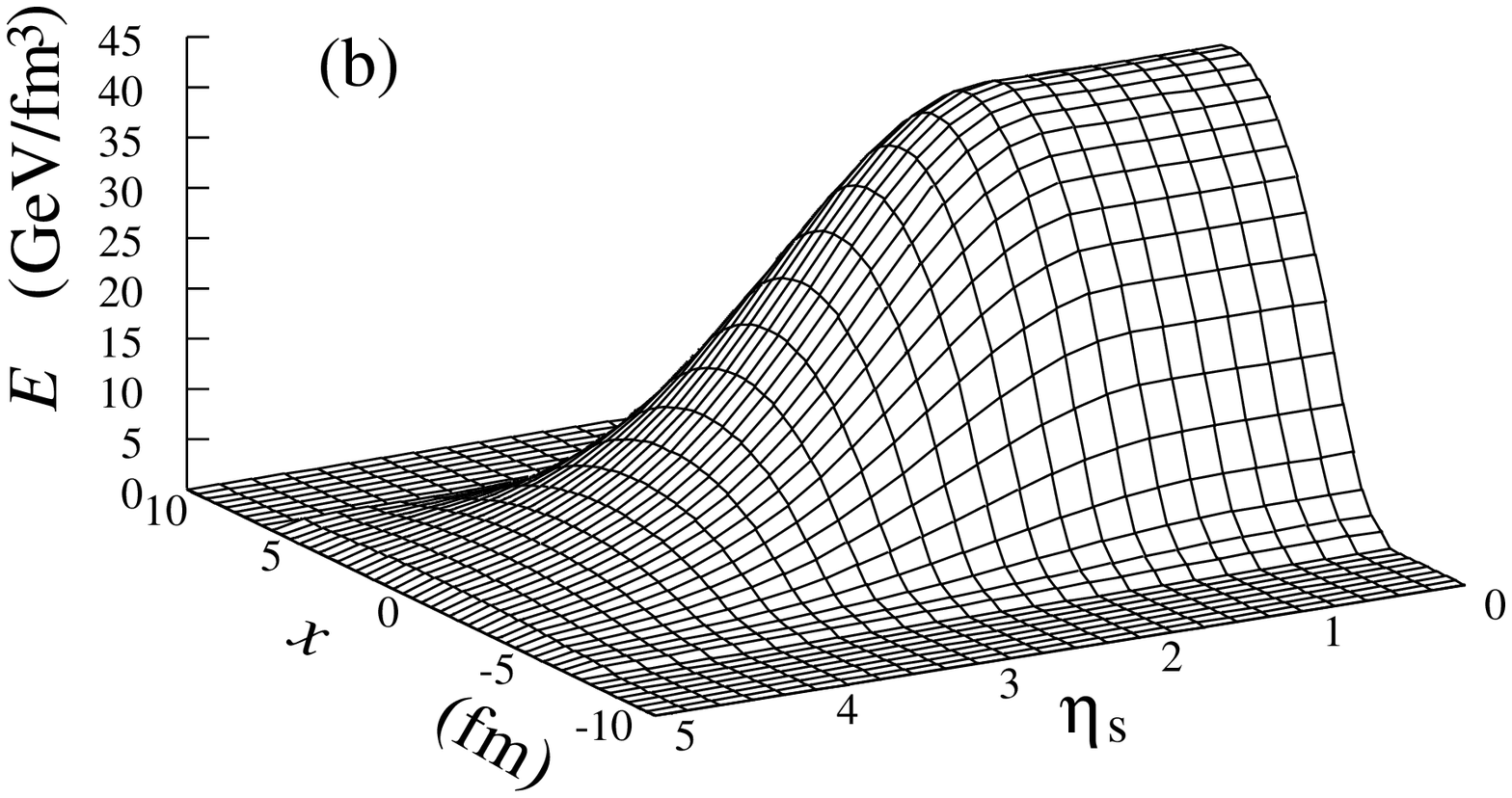,height=3.4cm}
  \end{minipage}
  \caption{\label{E_tau2} Energy density in the reaction plane of 
  central collisions obtained in the QGSM for $\tau=1~{\rm fm/c}$ -- left 
  panel (the same as on the left of Fig. \ref{E_tau1} but rescaled), 
  and initial distribution used in hydrodynamical calculations 
  by Hirano et al. -- right panel. The figure on the right is taken from Ref. 
  \cite{hydro}.}
\end{figure}
The direct comparison of the microscopically determined density profile 
on the one side and a hydrodynamical model assumption from Ref. \cite{hydro} 
on the other side -- illustrated in Fig. \ref{E_tau2} -- exemplifies this 
fact. 
Also in the transversal plane the shape and order of magnitude of 
$\epsilon$ at $\tau = 1~{\rm fm/c}$ is supported by a hydrodynamical 
calculation of Ref. \cite{hydro2}, where the NeXuS event generator 
\cite{NEXUS} has been used to determine the initial conditions for 
central Au+Au collisions at $\sqrt{s_{NN}} = 130~{\rm GeV}$. 
There it has been shown that fluctuations of the initial conditions 
produced by a microscopic model and used as input for hydrodynamical 
calculations are important and lead to discrepancies for some observables. 
Together with our findings for the averaged distributions of energy density 
the assumption of smooth and symmetric initial conditions seems to be a 
rather rough one.

\section{Local equilibration}
\label{eq}

As one can expect from the energy density distributions of Fig. \ref{FIG1}, 
global equilibrium cannot be reached; the distributions of particles are far 
from being isotropic for all time steps under consideration. 
But when one looks in more detail, some cells could reach local equilibrium. 
In order to measure the degree of kinetic equilibration for a given cell, 
the pressure components for the x-, y- and z-direction are calculated 
\cite{URQMD_edens}. 
The diagonal elements of the pressure tensor in a cell according to the 
virial theorem are given by \cite{pressure}
\begin{equation}
P_{\{x,y,z\}} = \frac{1}{3V}\,\sum_i 
\frac{p_{i,\{x,y,z\}}^2}{\sqrt{m_i^2 + p_{i,x}^2 + p_{i,y}^2 + p_{i,z}^2}}~,
\label{press_comp}
\end{equation}
where $V$ is the volume of the cell, $m_i$ is the mass and $p_{i\{x,y,z\}}$ 
are the momentum components of hadron $i~$. The sum runs over all hadrons 
in the corresponding cell. The pressure is locally evaluated, therefore 
a Lorentz transformation with the average velocity in the x-, y- and 
z-direction is applied before taking the sum. 
The requirement of isotropic velocity distributions, i.e. kinetic 
equilibrium, is closely related to the requirement of the pressure 
isotropy, which would imply $P_x = P_y = P_z~$. 
Therefore, the equilibration ratio $R_{\rm LE}$ as defined by
\begin{equation}
R_{\rm LE} = \frac{1}{2}\,\frac{P_x + P_y}{P_z}
\label{equi_ratio}
\end{equation}
is evaluated for each cell as a measure of local equilibration. 
Local equilibrium in the corresponding cell is then characterized by 
$R_{\rm LE}\rightarrow 1~$.

First of all, the equilibration process is studied for the central cell 
of the collision, given by $-2~{\rm fm} < x, y, z < 2~{\rm fm}$ for 
$b=0~{\rm fm}$ and 
$2~{\rm fm} < x < 6~{\rm fm}$, $-2~{\rm fm} < y, z < 2~{\rm fm}$ for 
$b=8~{\rm fm}$, by looking at the pressure for all three directions. 
\begin{figure}[ht]
  \centering
  \epsfig{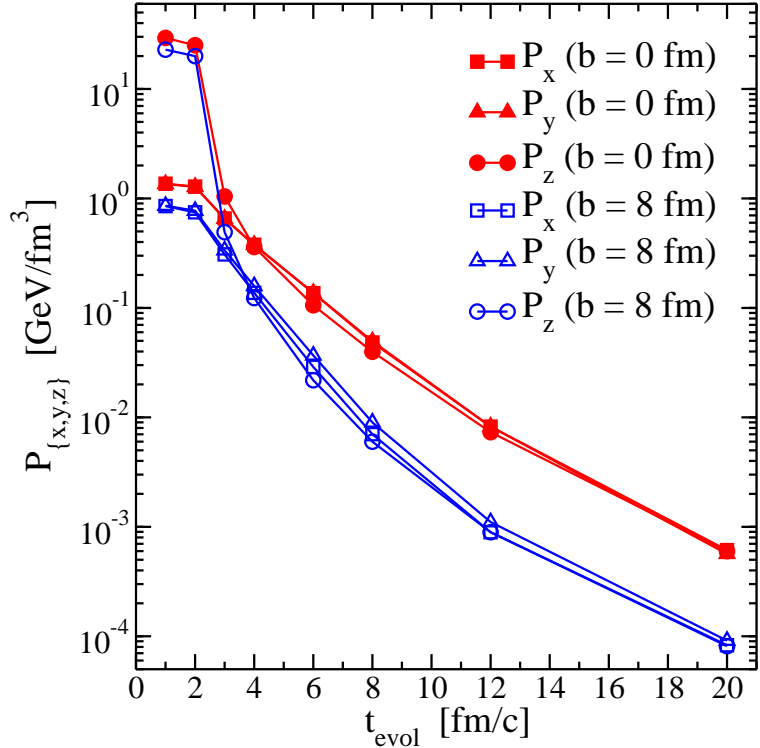}
  \caption{\label{EQ}Pressure components in x-, y- and z-direction in 
  the central cell of the overlap zone of the colliding gold nuclei with 
  $b = 0~{\rm fm}$ and $b = 8~{\rm fm}$, respectively.}
\end{figure}
Fig. \ref{EQ} shows the pressure components as defined in 
Eq. (\ref{press_comp}) for the central and the semi-peripheral collisions. 
In both reactions the longitudinal pressure is clearly dominating the very 
early evolution of the system while the transversal pressure components in 
x- and y-direction are nearly equal. 
At $t \approx 2~{\rm fm/c}$ the longitudinal pressure starts to 
decrease drastically and reaches a level close to the transversal pressure 
after $\approx 4~{\rm fm/c}$ time of evolution. Thereafter all the three 
corresponding pressure components decrease with similar slopes. 
Nevertheless, the pressure isotropy is not yet achieved because deviations 
up to $30-50~\%$ between the longitudinal and transversal components 
occur during the next $8~{\rm fm/c}~$. Following the discussion in 
\cite{URQMD_edens}, one can call this a pre-equilibrium stage . It sets in 
very quickly for both, the central and the semi-peripheral collision, 
whereas the dynamics of the kinetic equilibration process during the 
pre-equilibrium stage is somewhat different for both reactions.

\begin{figure}
  \centering
  \epsfig{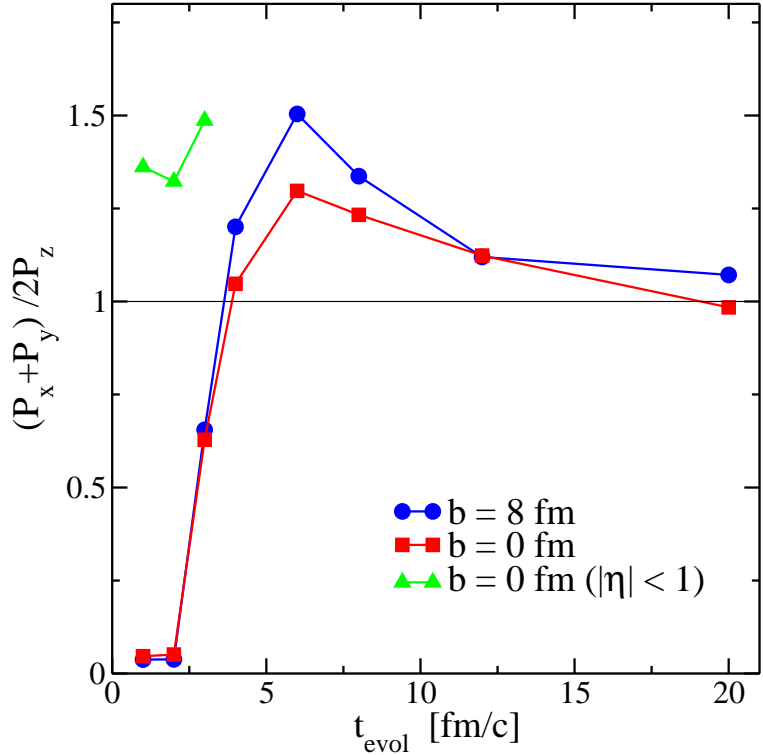}
  \caption{\label{EQ1}Degree of equilibration as defined by the 
  equilibration ratio $R_{\rm LE}$ in the central cell of the overlap zone 
  for Au+Au collisions at $\sqrt{s_{NN}} = 200~{\rm GeV}$ with 
  impact parameters $b = 0~{\rm fm}$ and $b = 8~{\rm fm}$. 
  The triangles indicate the equilibration ratio of particles with 
  pseudorapidities $|\eta| \le 1$ only, produced within the first 
  $3~{\rm fm/c}$ of the central collisions (see detailed discussion below).}
\end{figure}

This difference in the degree of equilibration is clearly seen, when one 
additionally looks at the equilibration ratio $R_{\rm LE}$ defined in 
Eq. \ref{equi_ratio}. Fig. \ref{EQ1} depicts $R_{\rm LE}$ in the central 
cell of the overlap zone for central and semi-peripheral Au+Au collisions. 
Again, the pre-equilibrium stage with $R_{LE} \approx 1$ is reached for 
central collisions after $4~{\rm fm/c}$ corresponding to a time 
when the high density areas as seen in Fig. \ref{FIG1} and Fig. \ref{FIG2} 
have completely passed through the central cell. Within these high density 
areas the longitudinal pressure is clearly dominating and therefore no 
equilibrium can be reached. During the next few fm/c the transverse 
pressure components are slightly dominating the expansion; the equilibration 
ratio tends to be greater than one. But it decreases already after 
$\approx 6~{\rm fm/c}$ time of evolution and -- as the cell is loosing 
more and more particles -- approaches almost local equilibrium after 
$12-15~{\rm fm/c}$. The transverse pressure components $P_x$ and 
$P_y$ are nearly equal to each other during this stage, what reflects 
the symmetry of the collision zone in the transverse plane.

In case of the semi-peripheral collision with an impact parameter of 
$8~{\rm fm}$ local equilibrium seems not to be reached in the cell 
corresponding to the overlap zone of the collision after $15~{\rm fm/c}$, 
but nevertheless $R_{\rm LE}$ gets at least very close to unity. After the 
high density areas passed through the cell, the transverse pressure is 
clearly dominating the next few fm/c of the pre-equilibrium stage -- 
stronger than in the central collision, but in contrast to the central 
reaction it is itself not equilibrated. As can be seen in Fig. \ref{EQ}, 
the y-component of the pressure tends to be greater (up to 30\%) than the 
x-component . This asymmetry arises due to the elliptical overlap of the 
colliding nuclei and reflects the smaller pressure gradients in the 
y-direction, which, in the hydrodynamical picture, is responsible for the 
build-up of elliptic flow \cite{hydro_flow}.

The question, to what extent the produced matter reaches thermalization and 
how fast this may be achieved, is still open. In Ref. \cite{bahl05} it was 
shown that the linear rise of elliptic flow, in particular $v_2/\varepsilon$ 
with $\varepsilon$ being the eccentricity of the spatial overlap of the 
colliding nuclei at RHIC, with the density gives rise to the conclusion 
of a density dependence and hence to a small number of collisions per 
particle. Thermalization could therefore not be reached so fast. 
Our findings of equilibration at around $t \approx 15~{\rm fm/c}$ together 
with the elliptic flow results and their time evolution obtained in the QGSM 
\cite{QGSM_flow2,QGSM_flow3,QGSM_flow4} support this picture and contradict 
quasi the usual hydrodynamical assumptions of fast thermal equilibration.

However, as can be seen in Fig. \ref{EQ1} a certain fraction of the produced 
particles, i.e. hadrons with pseudorapidities $|\eta| \le 1$, is closer to 
local equilibrium even at times $t\approx 1~{\rm fm/c}$. In between the two 
currents of matter of the target and projectile fragmentation regions, which 
make up $\approx 2/3$ of the energy density, this fraction provides a dense 
and roughly equilibrated local surrounding even at very early times of central 
heavy ion collisions. 
Thus we find energy densities of about $4-5~{\rm GeV~fm}^{-3}$ in central 
Au+Au reactions at $\sqrt{s_{NN}} = 200~{\rm GeV}$ for matter which can be 
considered to be at least in a pre-equilibrium stage already at 
$1-2~{\rm fm/c}$. The preconditions for a phase transition, i.e. energy 
densities $\epsilon > \epsilon_{\rm crit}\simeq 1 ~{\rm GeV~fm}^{-3}$, 
are given for a duration of $\approx 8~{\rm fm/c}$ in a space-time volume 
of about $500~{\rm fm}^{4}$.

\begin{figure}[htb]
  \psfig{file=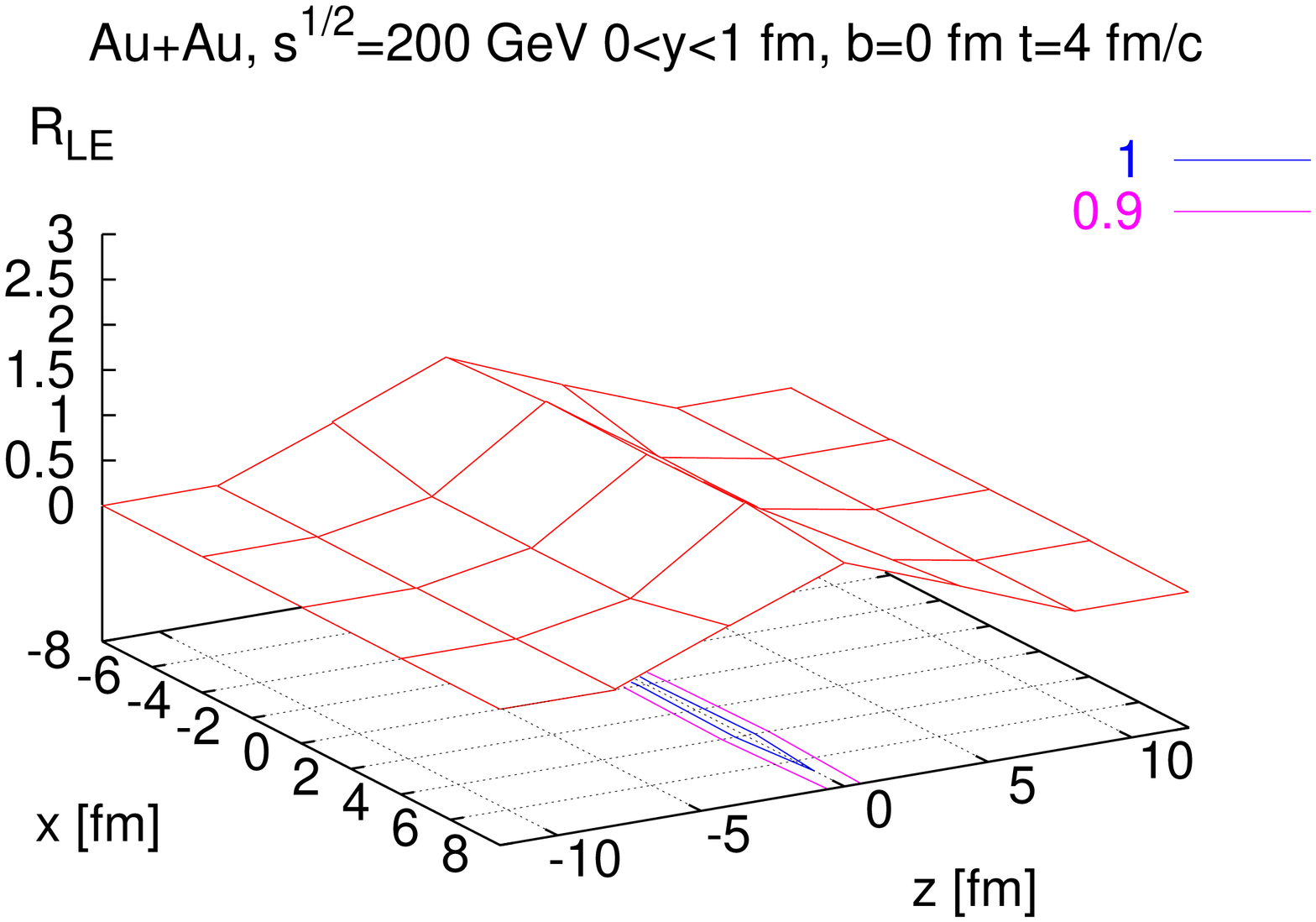,height=6.8cm,angle=-90}
  \psfig{file=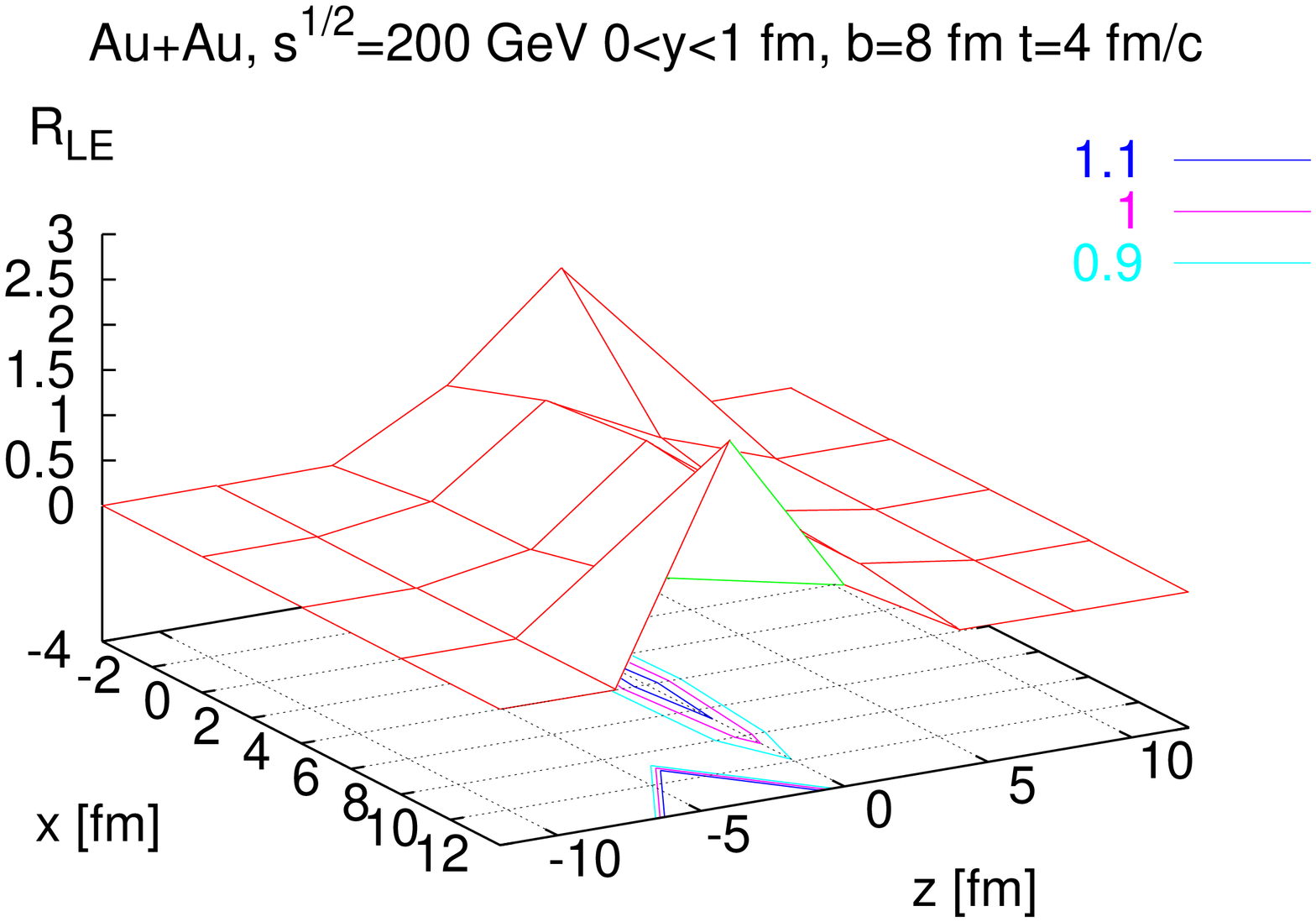,height=6.8cm,angle=-90}
  \psfig{file=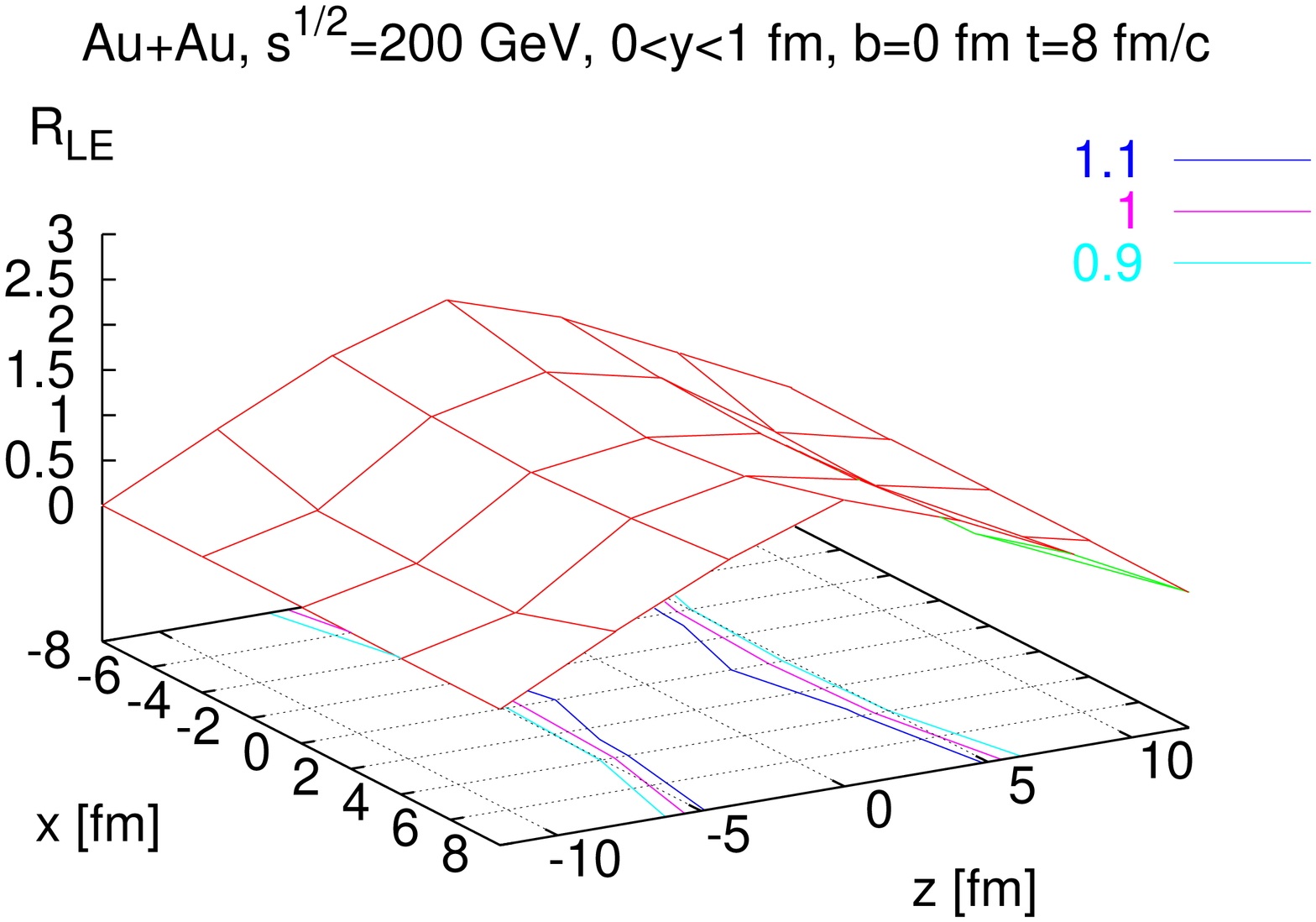,height=6.8cm,angle=-90}
  \psfig{file=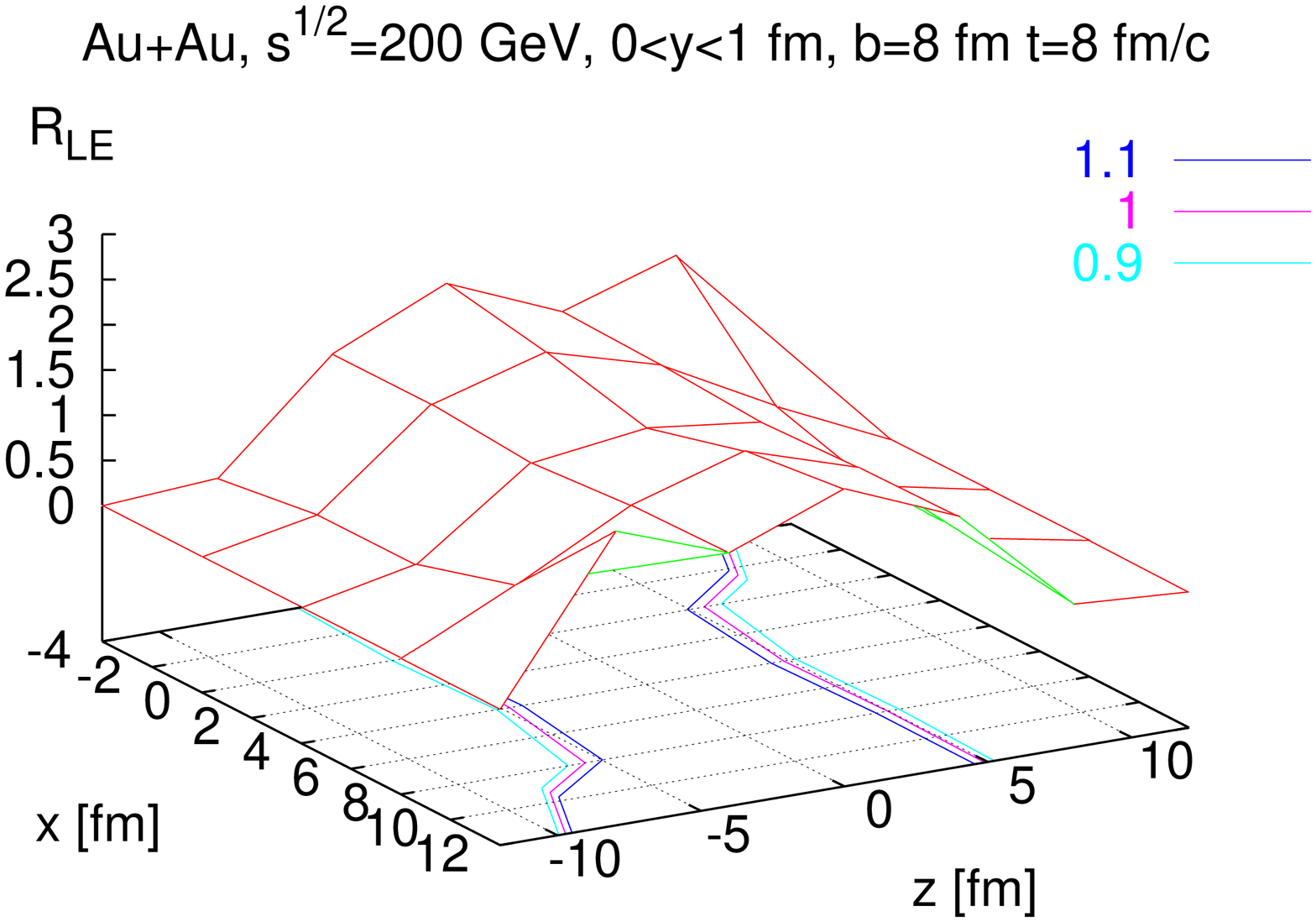,height=6.8cm,angle=-90}
  \psfig{file=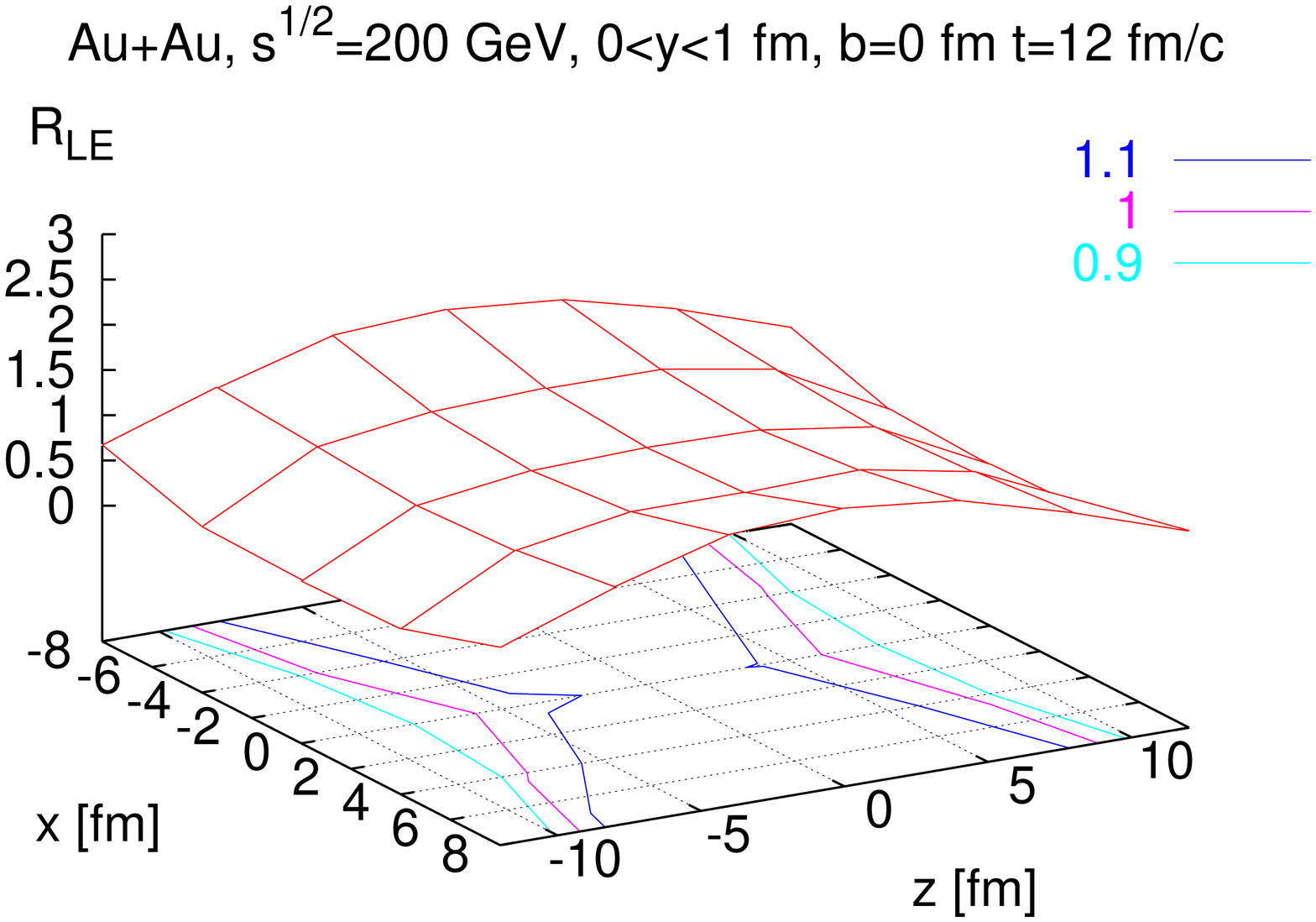,height=6.8cm,angle=-90}
  \psfig{file=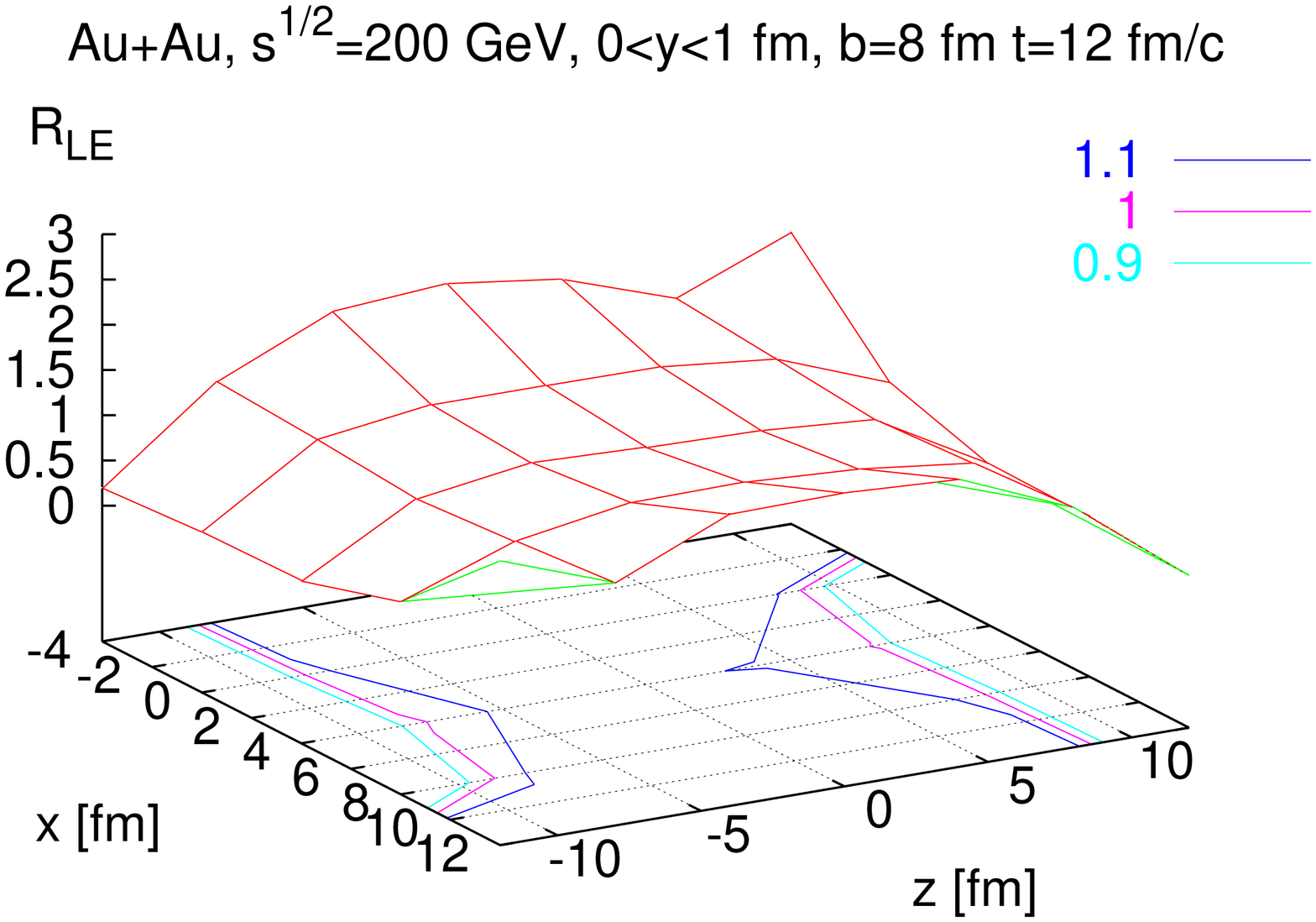,height=6.8cm,angle=-90}
  \caption{\label{EQ4fm}Snapshots of the equilibration ratio $R_{LE}$ 
  in the overall x-z plane after $t=4,~8,~{\rm and}~12~{\rm fm/c}$ time of 
  evolution. The distributions on the left are for collisions with 
  $b=0~{\rm fm}$, and on the right for collisions with $b=8~{\rm fm}$.}
\end{figure}

Finally, the degree of local equilibration and its time evolution is evaluated 
in the whole reaction ($x-z$) plane for the central and semi-peripheral Au+Au 
collisions. All the results depicted in Fig. \ref{EQ4fm} for the three time 
steps $t = 4,~8,~{\rm and}~12~{\rm fm/c}$ show a characteristic saddle 
structure, which is more smoothly and faster developed in the central than 
in the semi-peripheral reaction. 
For the x-direction the equilibration ratio $R_{LE}$ always has a minimum 
in the overlap zone whereas it reaches here the maximum as a function of 
the longitudinal coordinate z. The high density areas in front of the 
expanding system, in other words at large values of z, never reach local 
equilibrium; only in the aftermath of the pass-through the inner cells tend 
towards kinetic equilibration indicated by $R_{LE} \approx 1$. 
In the center of the collision the transverse pressure is dominating at all 
times, therefore is $R_{LE} > 1$. But towards higher densities in the 
z-direction the longitudinal, i.e. z-component of the pressure is increasing 
and therefore supplying more equilibrated zones filled with outward moving 
particles. Since the most collisions naturally happen within the high density 
areas, it is important to notice that the most secondary particle interactions 
take place in a non-equilibrated surrounding.

Closing the discussion, it should be mentioned that string-cascade models 
provide most likely an upper limit for relaxation times towards local 
equilibrium. This is not only due to the fact that models of this type 
do not contain a phase transition to partonic degrees of freedom which 
are expected to equilibrate much faster. Standard string-cascade models 
violate detailed balance since particle production through string decays 
leads to processes $2 \longrightarrow N$ with two initial and N final 
state hadrons. However, corresponding backward processes 
$N \longrightarrow 2$ are not included. 
This approximation is usually justified by the assumption that many-body 
scattering processes $N\longrightarrow X$ play a minor role in open, fast 
expanding systems. However, when high particle densities are achieved 
such processes can become relevant, e.g. for the production of rare hadron 
species \cite{rapp01}. Many-body processes, such as string or hadron fusion 
and recombination, lead naturally to shorter equilibration times, as has 
e.g. be demonstrated in \cite{greiner05} for the gluonic sector. 
In very dense states of matter it is expected that interactions among 
strings, before they fragment into hadrons, are able to accelerate the 
equilibration process in microscopic string models. Possible concepts in 
this context are the before mentioned string fusion, the formation of string 
clusters, and eventually percolation of strings, which have been considered 
in several string-cascade models \cite{Sorge:1995dp,Vance:2000ax,Bratkovskaya:2004kv,Amelin:2001sk,Pajares:2005kk}. 
Effectively, such processes describe reactions of $2n~ (n \ge 2)$ initial 
particles going to $N$ final state hadrons. Nevertheless, detailed balance 
is still violated. But future studies should be dedicated to the question 
how far these multi-string processes are able to reduce quantitatively the 
relaxation times towards local equilibrium.

\section{Summary and conclusions}

In this survey, we have analysed several aspects of the collision dynamics 
in central and semi-peripheral Au+Au reactions at the highest RHIC center 
of mass energy of $\sqrt{s_{NN}}=200~{\rm GeV}$ within the microscopic 
quark-gluon string model (QGSM), which is based on the colour exchange 
mechanism for string formation.

The first major topic was dedicated to the time evolution of the local energy 
density $\epsilon$. Maximum energy densities of around $15~{\rm GeV/fm^3}$ 
for the central collisions and around $10~{\rm GeV/fm^3}$ in semi-peripheral 
reactions are reached in the overlap zone after the first initial hadron 
collisions. The energy density stays almost constant in the central cell 
during the first $2~{\rm fm/c}$ time of evolution. The value of $\epsilon$ 
in this cell is rapidly decreasing as soon as the fastest particles start 
to leave the cell. This happens at around $t\approx 2~{\rm fm/c}$. 
The corresponding $\epsilon$ distributions in the x-z plane as a function 
of the system time $t$ in the c.m. frame of the colliding nuclei reveal, 
that most particles do not stay in the central collision zone but form 
two disks in the fragmentation regions occupied with large numbers of 
produced particles which travel practically on the light cone in beam 
$(z)$ direction. Also as a function of the boost-invariant space-time 
rapidity $\eta_s$, the system has an energy density distribution with a 
nontrivial shape in the reaction plane, i.e. a characteristic double-peaked 
structure in $\eta_s$-direction which is maintained in the central and also 
in the semi-peripheral collisions. Complementary to the energy density in the 
reaction plane, the corresponding distributions of $\epsilon$ in the 
transversal plane and additionally at constant proper time $\tau=1~{\rm fm/c}$ 
were studied (only for the central reaction). The QGSM results of the 
latter distributions compare well with hydrodynamical assumptions for 
initial energy density profiles \cite{hydro,hydro1,hydro2}. These are 
similar in shape as well as in the order of magnitude. 
Nevertheless, the absolute values of $\epsilon$ drop rapidly when the system 
expands and the peaks get significantly washed out due to ongoing secondary 
collisions between the produced particles. But we want to emphasize that, 
although $\epsilon$ is rapidly decreasing in the transverse directions 
as well as in time, there definitely exist regions with local energy 
densities higher than the expected critical energy density for the QCD 
phase transition within the first $8~{\rm fm/c}$.

However, at that time the system is not completely in local equilibrium, 
as it was elaborated in the second major topic. There the question whether 
and how fast the produced particles in ultra-relativistic Au+Au collisions 
equilibrate has been studied. 
It was shown that in both reactions, i.e. central and semi-peripheral ones, 
the longitudinal pressure is clearly dominating within the first 
$2~{\rm fm/c}$. After the fast particles have left the central cell, all 
the three pressure components converge to each other. The next $8~{\rm fm/c}$ 
the transverse pressure is slightly stronger, not allowing the system to 
reach full local equilibrium but entering some kind of pre-equilibrium stage. 
As the central cell of the overlap zone is loosing more and more particles, 
it approaches almost local equilibrium after $12-15~{\rm fm/c}$ in the central 
collisions and $15-20~{\rm fm/c}$ in case of the semi-peripheral reactions. 
Finally, the degree of local equilibration characterized by $R_{LE}$ and its 
time evolution in the whole reaction ($x-z$) plane shows in the central and 
semi-peripheral Au+Au collisions for all considered time steps a 
characteristic saddle structure, which indicates that the high density 
areas in front of the expanding system are far away from local equilibrium. 
Only the inner cells tend towards kinetic equilibration. Since the most 
secondary collisions happen within these high density areas, the most 
particle interactions take place in a non-equilibrated surrounding. 
Global equilibrium is not reached at all -- the global particle distributions 
are far from being isotropic for all time steps under consideration.

As it was already mentioned, the question to what extent the produced matter 
reaches thermalization and how fast this may be achieved in the experiment 
is still open. Our findings of almost full local equilibration at around 
$t \approx 15~{\rm fm/c}$ together with the elliptic flow results and their 
time evolution obtained in the QGSM \cite{QGSM_flow2,QGSM_flow3,QGSM_flow4} 
support the idea that fast thermal equilibration is not necessary to 
produce large elliptic flow, what contradicts somehow the usual hydrodynamical 
assumptions. Nevertheless, a pre-equilibrium stage with energy 
densities well above the critical energy density is already 
achieved at early time scales. This means that the preconditions 
for a phase transition can be found within the string-cascade approach 
which, due to the lack of multi-string and/or multi-particle fusion processes, 
provides an upper limit for the equilibration times.

In this context, it is also important to stress, that all calculations are 
performed by averaging over 200 and 600 events for central and semi-peripheral 
reactions respectively. For a single collision event the energy density 
as well as local equilibrium is subject to strong fluctuations. It is 
not a priori clear to what extent these fluctuations may contribute to 
the final observables, as has been pointed out in Ref. \cite{hydro2}. 
This question together with a more detailed understanding of the freeze-out 
scenario for the hadron production and the influence of multi-string/particle 
reactions on equilibration deserves further investigations.

\section*{Acknowledgments}

This work was supported by the Bundesministerium f{\"u}r Bildung und 
Forschung (BMBF) under contract 06T\"U986.



\end{document}